\documentclass[%
 reprint,
nofootinbib,
 amsmath,amssymb,
 aps,
prstab,
]{revtex4-2}
\usepackage[caption=false]{subfig}
\usepackage{diagbox}

\usepackage{graphicx}
\usepackage{dcolumn}
\usepackage{bm}

\numberwithin{equation}{section}

\begin{document}

\preprint{APS/123-QED}

\title{Creating Exact Multipolar Fields with Azimuthally Modulated RF Cavities}%

\author{L. M. Wroe}
 \email{laurence.wroe@physics.ox.ac.uk}
\author{S. L. Sheehy}%
\altaffiliation[Also at ]{School of Physics, University of Melbourne.}
\affiliation{%
 Department of Physics, University of Oxford, OX1 3RH, United Kingdom
}%


\author{R. J. Apsimon}
\affiliation{
 Engineering Department, Lancaster University, LA1 4YW, United Kingdom, and \\Cockcroft Institute,
Daresbury Laboratory, Warrington, WA4 4AD, United Kingdom}

\date{\today}

\begin{abstract}
RF cavities used in modern particle accelerators operate in  TM$_{m10}$-like modes composed of a single, dominant multipole of order $m$;  $m=0$ modes are used for the longitudinal acceleration of a particle beam and $m\neq0$ modes for controlling transverse beam dynamics. The practical design of the latter, however, can be complex and require extensive analysis through the iteration of both approximate mathematical models and computationally expensive simulations to optimise the performance of the structure. In this paper we present a new, systematic method for designing azimuthally modulated RF cavities that support modes composed of any number and magnitude of user-specified transverse multipoles, either with or without a longitudinally accelerating component. Two case studies are presented of RF cavity designs that support modes composed of a longitudinally accelerating field  in addition to a single transverse multipole, and designs that support modes composed of two transverse multipoles. We discuss generalising the discoveries and conclusions from the two case studies to designing cavities that support modes composed of any number of multipoles. The theoretical work is verified with analysis of 3D simulations and  experimental measurements are presented of a cavity operating in a 3 GHz mode that simultaneously longitudinally accelerates  and transversely focuses a beam.

\end{abstract}

\maketitle

\section{\label{sec:level1}Introduction}

RF cavities that control transverse beam dynamics by operating in transverse magnetic TM$_{m10}$-like modes composed of a single, dominant transverse multipole of order $m\neq0$ have a number of applications  in particle accelerators. For example, cavities operating in a TM$_{110}$-like mode with a dominant dipole component in the field have application in: separating a single particle beam into multiple beams \cite{Deflector1, Deflector2, Deflector3}, providing longitudinal diagnostic measurements on a particle beam such as measuring its emittance \cite{DeflectingDiagnostics}, controlling electron beams in free-electron lasers by acting as emittance exchangers \cite{EmittanceExchange2}, compressing x-ray pulses in synchrotron light sources \cite{PulseComp}, and providing a head-tail rotation to particle bunches to allow for luminosity control in particle colliders as crab cavities \cite{Palmer}. The latter were investigated at KEKB \cite{KEKB} with several designs \cite{Crab_Cavities} proposed and developed for the HL-LHC \cite{HLLHC}. Additionally, cavities operating in a TM$_{210}$-like mode with a dominant quadrupolar component are being investigated for application in Landau damping transverse beam instabilities in the HL-LHC \cite{CERN_RFQ}. 

With $m$, $n$, and $p$ respectively denoting the azimuthal, radial and longitudinal order of the electromagnetic mode, it is a known and well-studied result   that a pillbox cavity, that is an enclosed RF cavity with a circular cross-section, will support TM$_{mnp}$-exact modes with unique resonant frequencies for distinct values of $m$, $n$, and $p$ \cite{Lee}. In order to use an RF cavity that operates in a $m\neq0$ TM$_{mnp}$ mode in a particle accelerator, novel and elaborate designs \cite{DeflectingDiagnostics, DampingModes, 4RCC} that introduce azimuthal and/or longitudinal asymmetries are required to: separate out any degenerate TM$_{mnp}$ same-order modes in frequency space, damp any lower-, same-, and higher-order modes that interfere significantly with beam dynamics, and to optimise the figures of merit such as R/Q. Due to the complexity of such designs, the exact mathematical form of the EM field of the desired mode, and therefore its exact effect on beam dynamics, cannot be solved for analytically. As a result, computationally expensive simulations are required to optimise designs \cite{TaylorMaps, 4RCC}, and these may be used in tandem with  novel mathematical models developed specifically to provide quick but approximate insights into the  optimisation process \cite{EquivalentCircuit}. What's more, the implementation of azimuthal and longitudinal asymmetries introduces unwanted transverse multipoles into the desired TM$_{mnp}$ mode, and analysis is required to ensure these unwanted multipoles do not exceed tolerance levels for the accelerator's operation \cite{LongTermDynamics}. If tolerance levels are exceeded, then this may require a redesign of the RF cavity system: for example, RF cavities that operate in TM$_{010}$-like modes can be designed with dual-port power couplers instead of single-port couplers to prevent the introduction of unwanted dipolar components \cite{PortCoupler}.

In addition to the application of RF cavities that operate in modes composed of a single multipole, designing RF cavities to operate in modes composed of numerous multipoles could have useful application in particle accelerators. For example, the use of elliptical irises  to introduce transverse quadrupolar focusing in accelerating RF cavities was investigated \cite{CLIC_Note_34, CLIC_Note_62,CLIC_Note_143} for use in the proposed future linear collider CLIC \cite{CLIC_CDR}.

In this paper, we present a systematic method for designing RF cavities that support modes composed of any desired number and magnitude of multipolar components. We define these modes with the notation TM$_{\{M\}\eta p}$. $\{M\}$ denotes the subset of integers representing the multipoles composing the mode, for example the TM$_{\{0,1\}\eta p}$ modes are solely composed of a longitudinally accelerating monopolar component and a transverse dipolar component. We use $\eta$ to denote and distinguish the more complex radial order of general TM$_{\{M\}\eta p}$ modes from the simpler TM$_{mnp}$ modes, and $p$ continues to denote the longitudinal order of the mode. The method can be applied to modes with any value of $p$. For simplicity, however, we study the subset of $p$ = 0 modes with no longitudinal variation because the vast majority of RF cavities in particle accelerators operate in $p=0$ modes. The RF cavities that support TM$_{\{M\}\eta 0}$ modes containing at least two distinct, multipolar components are termed \textit{azimuthally modulated cavities} because we find that the cavity cross-section, $r_0(\theta)$, must vary with the azimuth. Such cavities have benefits and application in particle accelerators because the TM$_{\{M\}\eta 0}$ mode: can exactly introduce any number and magnitude of desired multipolar components, is non-degenerate which overcomes the need to separate out a same-order mode in frequency space, and has a field profile  that can be expressed analytically throughout the cavity. 

Section \ref{sec:Theory} begins with the derivation of the transcendental equation that must be numerically solved to determine the azimuthally modulated cross-sections that support specific TM$_{\{M\}\eta 0}$ modes. Insight into the properties of these azimuthally modulated cavity shapes can be gained using mathematical and graphical methods and we first introduce these techniques by applying them to the simple case of the TM$_{mn0}$ modes, known to be supported by cavities with a pillbox shape. We then present two case studies of the azimuthally modulated RF cavity shapes that support the TM$_{\{0,m_1\}\eta 0}$ modes containing a longitudinally accelerating monopole and a single transverse multipole, and those that support the TM$_{\{m_1,m_2\}\eta 0}$ modes containing two transverse multipoles. We afterwards discuss generalising the discoveries and conclusions made in these two case studies for the design of cavity shapes that support modes with any number of multipoles. In Sec. \ref{sec:CSTSimulations}, we present and analyse simulation results of azimuthally modulated RF cavities designed to support the TM$_{\{0,3\}\eta0}$ modes containing a monopolar and a sextupolar component. Finally, in Sec. \ref{sec:QuadBuild} we show the results of experimental bead pull measurements of a novel RF cavity operating in a 3 GHz TM$_{\{0,2\}20}$ mode.

\section{\label{sec:Theory}Theory}
\subsection{Deriving the Form of a TM$_{\{M\}\eta0}$ Mode}\label{sec:TheoryHelm}

In this paper, we study perfectly conducting, enclosed RF cavities of longitudinal extent $L$ and cross-section $r_0(\theta)$. The modes supported by these azimuthally modulated cavities must have an electric field that satisfies the boundary conditions $E_r(r,\theta,\pm L/2) = E_\theta(r,\theta,\pm L/2) = 0$ at the cavity endpoints and $E_z(r_0(\theta),\theta,z) = 0$ at the cavity surface. Imposing the former boundary condition and assuming that the particle's transverse momentum  is negligible compared to its longitudinal momentum, Panofsky and Wenzel derive that the change in transverse momentum of a particle that passes through an RF cavity can be determined solely by the transverse gradient of the longitudinal component of the electric field, $E_z(r,\theta,z)$ \cite{PanofskyWenzel1953}:
\begin{equation} \label{eq:PWEqn}
    \Delta \vec{p}_\perp(r,\theta) =  -i\frac{q}{\omega}\int_{-L/2}^{L/2} \nabla_\perp E_z(r,\theta,z) dz,
\end{equation}
\noindent where $q$ is the charge of the particle and $\omega$ is the resonant angular frequency of the RF cavity mode. The inclusion of the imaginary unit means the transverse force is $\pi/2$ out of phase with the longitudinal force. 

The Panofsky-Wenzel theorem therefore states that only the longitudinal electric field of the mode need be determined in order to calculate the effect of the transverse forces that act on a particle beam passing through a cavity. The general form of the electric field, $\vec{E}(r,\theta,z)$, of a mode supported by an RF cavity of any design is derived in Ref. \cite{Abell2006}. This general form can be simplified to determine the specific form of the longitudinal electric field in azimuthally modulated cavities by applying the boundary conditions at the endpoints and cavity surface. Applying the endpoint boundary condition, we find that modes of distinct longitudinal order $p$ must independently satisfy the boundary conditions. Applying the surface boundary condition, we find that the allowed modes belong to one of two familiar sets: either TM modes with no longitudinal magnetic field or TE modes with no longitudinal electric field. In this paper, we only study TM modes with $p=0$: TE modes are ignored as they are less frequently used in particle accelerators and their effect on beam dynamics cannot be analysed using the Panofsky-Wenzel theorem.

The general form of the longitudinal electric field given in Ref. \cite{Abell2006} reduces in the case of TM$_{\{M\}\eta 0}$ modes to:
\begin{equation} \label{eq:EzForm}
    E_z(r,\theta,z) = J_0(kr)\tilde{g}_0+ \sum_{m=1}^\infty
    J_m\left(kr\right)\tilde{g}_{m}\cos{(m\theta-\phi_m)} ,
\end{equation}
\noindent where $k=\omega/c$ is the wavenumber of the mode, $\tilde{g}_{m}$ and $\phi_m$ are arbitrary constants that respectively denote the magnitude and offset of the multipole of azimuthal order $m$,  and $J_m$ denotes the Bessel function of the first kind of order $m$. We note that $\phi_m=0$ corresponds to a normal multipolar field and $\phi_m = \pi/2$ to a skew multipolar field.

In this paper, the aim is to determine a method for designing RF cavities that support desired modes composed of multipoles of specified magnitudes. If we consider the set of azimuthally modulated cavities with cross-sections $r_0^{(\eta)}(\theta)$ that support a desired TM$_{\{{M}\}\eta 0}$ mode, then the boundary condition $E_z(r_0^{(\eta)}(\theta),\theta,z)=0$ must be satisfied at the surface of each of these cavities. Applying this to Eq. \ref{eq:EzForm} gives:
\begin{equation} \label{eq:EzToSolve} 
    J_0\!\left(kr_0^{(\eta)}(\theta)\right)\!\tilde{g}_0 + \sum_{m=1}^\infty\!
    J_m\!\left(kr_0^{(\eta)}(\theta)\right)\!\tilde{g}_{m}\cos{(m\theta-\phi_m)}\!=0,
\end{equation}
\noindent Equation \ref{eq:EzToSolve} is transcendental and must, in general, be solved  using numerical methods to determine $r_0^{(\eta)}(\theta)$. Doing so for user-specified values of $k$, $\tilde{g}_m$ and $\phi_m$, returns the set of unique azimuthally modulated cross-sections, $r_0^{(\eta)}(\theta)$, that support the corresponding TM$_{\{{M}\}\eta 0}$ modes.

\subsection{Modes with a Single Multipolar Component} \label{sec:IsolatingSingle}

Although numerical methods must generally be used to solve Eq. \ref{eq:EzToSolve} for all $\theta$, insight into the properties of the azimuthally modulated RF cavities that support TM$_{\{M\}\eta 0}$ modes can be gained through analysis using mathematical and graphical methods. 

To introduce the techniques used for understanding the properties of azimuthally modulated cavities that support modes with more than a single multipole, we briefly analyse the TM$_{m_1n0}$ modes that are known to be supported by a circular pillbox shape \cite{Lee}. For this case only, Eq. \ref{eq:EzToSolve} can be solved analytically:
\begin{equation}\label{eq:SingleMultipole}
    J_{m_1}\left(kr_0^{(n)}(\theta)\right) \tilde{g}_{m_1}\cos{(m_1\theta-\phi_{m_1})} = 0,
\end{equation}
is solved for all $\theta$ if:
\begin{equation} \label{eq:Bessel1Sol}
    J_{m_1}\left(kr_0^{(n)}(\theta)\right) = 0 \;\; \rightarrow \;\;  \frac{\omega r_0^{(n)}}{c} = j_{m_1n},
\end{equation}
\noindent where we define $\phi_0=0$ and $j_{m_1n}$ is the $n^{\textrm{th}}$ root of the Bessel function of order $m_1$. This shows that the cross-section of the cavity that supports a TM$_{m_1n0}$ mode is a circle of constant radius, $r_0^{(n)} = cj_{m_1n}/\omega$, as expected.

There are a number of important, known results for the TM$_{m_1n0}$ modes \cite{Lee}. Firstly, because all Bessel functions have $(n-1)$ turning points in the range [0, $j_{m_1n}$] and the field varies with the azimuth simply as $\cos{(m_1\theta-\phi_{m_1})}$, the integer $n$ is the number of poles (that is minima or maxima) of $E_z$ along any radial line that connects the centre of the cavity to its surface, except the nodal lines orientated at $\theta_q = (2q+1)\pi/2m_1+\phi_{m_1}$ ($q \in \mathbb{Z}$). Secondly, the $m_1\neq0$ modes are degenerate because the boundary condition in Eq. \ref{eq:SingleMultipole} is satisfied for all values of $\phi_{m_1}$. In contrast, the TM$_{0n0}$ modes are not degenerate. Thirdly, as $j_{m_1n}$ is unique for all distinct values of $m_1$ and $n$, the resonant frequencies of each of the TM$_{m_1n0}$ modes of distinct $m_1$ and $n$ are unique. Thus to use the TM$_{mn0}$ modes to introduce numerous multipoles at the same frequency, multiple pillbox cavities of different radii would have to be used. Finally, the TM$_{010}$ mode is the fundamental TM$_{mnp}$ mode with the lowest resonant frequency.

\begin{figure} 
\includegraphics[width=0.48\textwidth]{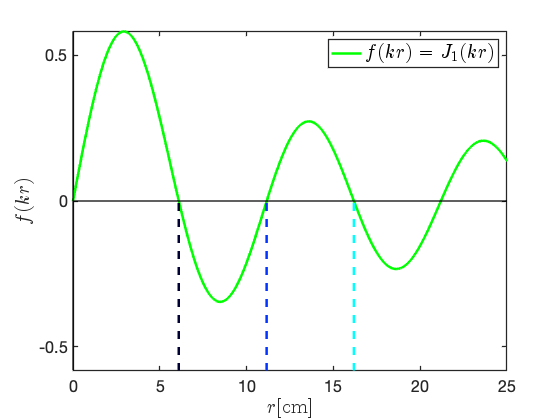} 
\caption{$J_1(kr)$ plotted against $r$ with $k=2\pi \times 3$ GHz/c. The black, blue, and cyan dashed lines show three points of intersection with the $r$-axis.} \label{fig:DipoleBessel}
\end{figure}

We can also use a graphical method to argue that TM$_{m_1n0}$ modes will be supported by cavities with a circular cross-section. Considering the 3 GHz TM$_{1n0}$ modes with $\tilde{g}_1 = 1$ and $\phi_1 = 0$, Eq. \ref{eq:SingleMultipole} states that the $r_0^{(n)}(\theta_0)$ solutions at a particular angle $\theta_0$ occur where, if plotted as a function of $r$, $J_1(kr)\cos{\theta_0}$ intersects the horizontal $r$-axis.  Figure \ref{fig:DipoleBessel} plots $J_1(kr)$ (the case where $\theta_0$ = 0) for $r=[0, 25]$ cm. The first three positive intersections  with the $r$-axis are indicated by the dashed black, blue, and cyan lines that correspond  respectively to the $r^{(n)}(0)$ solutions with $n$ = 1, 2, and 3. 

We could next consider plotting $J_1(kr)\cos{\theta}$ for many $\theta$ values and determining where the intersections lie for each $\theta$. This is not necessary, however, because the cosine term in $J_1(kr)\cos{\theta}$ acts as a $\theta$-dependent amplifying term on the Bessel function and thus the locations of the intersections with the $r$-axis will not change. As a result, we conclude that the $r_0^{(n)}(\theta)$ solutions are constant with respect to changes in $\theta$ and so plotting any $r_0^{(n)}(\theta)$ solution in Cartesian space plots a circle. This graphical argument can be generalised to argue that TM$_{m_1n0}$ modes of any $\tilde{g}_{m_1}$ and $\phi_{m_1}$ will be supported by a circular pillbox.

\subsection{Modes with a Monopolar and a Single Transverse  Multipolar Component} \label{sec:IsolatingSingleMono}

We now investigate the properties of the TM$_{\{0,m_1\}\eta 0}$ modes and the corresponding azimuthally modulated cross-sections that support them. We begin with general mathematical analysis. 

From Eq. \ref{eq:EzForm}, the subset of TM$_{\{0,m_1\}\eta 0}$ modes with $\phi_{m_1} = 0$  have a longitudinal electric field of the form:
\begin{equation}\label{eq:MonoSingleMultipoleField}
    E_z(r,\theta,z) = J_0\left(kr\right)\tilde{g}_0 + J_{m_1}\left(kr\right)\tilde{g}_{m_1}\cos{m_1\theta},
\end{equation}
and from Eq. \ref{eq:EzToSolve}, must satisfy the boundary condition:
\begin{equation}\label{eq:MonoSingleMultipole}
    J_0\left(kr_0^{(\eta)}(\theta)\right)\tilde{g}_0 + J_{m_1}\left(kr_0^{(\eta)}(\theta)\right)\tilde{g}_{m_1}\cos{m_1\theta} = 0.
\end{equation}

The periodicity of the $\cos{m_1\theta}$ function in Eq. \ref{eq:MonoSingleMultipole} means that both the longitudinal electric field of a TM$_{\{0,m_1\}\eta 0}$ mode and the corresponding azimuthally modulated cross-section have $m_1$-fold rotational symmetry. Additionally, $r_0^{(\eta)}(0)$ and $r_0^{(\eta)}(\pi/m_1)$ are turning points of opposite forms in the cross-section and so if $r_0^{(\eta)}(0)$ is a maximum then $r_0^{(\eta)}(\pi/m_1)$ is a minimum, or vice versa. 

Furthermore, if we consider the specific angles $\theta_q = (2q+1)\pi/(2m_1)$ $(q \in \mathbb{Z})$, the cosine term vanishes and Eq. \ref{eq:MonoSingleMultipole} becomes:
\begin{equation} \label{eq:MonoSingleFreq}
    J_0\bigg(kr_0^{(\eta)}\left(\theta_q\right)\bigg)\tilde{g}_0 = 0 \;\; \rightarrow \;\; \frac{\omega}{c}r_0^{(\eta)}\left(\theta_q\right) = j_{0x},
\end{equation}
\noindent where $x \in \mathbb{Z}$.\footnote{The reason $x\neq\eta$ is due to the possibility of conditional modes, discussed in Sec. \ref{sec:SingleMonoConditional}.} As well as relating the cavity frequency and radial value at these angles to the Bessel roots $j_{0 x}$, Eq. \ref{eq:MonoSingleFreq} defines $\eta$ for the TM$_{\{0,m_1\}\eta 0}$ modes as the number of poles in $E_z$ specifically along the $\theta_q = (2q+1)\pi/(2m_1)$ radial lines. This can be contrasted with the generality of the interpretation of $n$ as the number of poles in a TM$_{mnp}$ mode along any non-nodal, radial line.

As per the TM$_{0n0}$ modes in the circular pillbox cavity, the TM$_{\{0,m_1\}\eta 0}$ modes supported by azimuthally modulated cavities are also non-degenerate with no same-order mode.  Furthermore, by the Courant nodal domain theorem \cite{CourantNodal}, the TM$_{\{0,m_1\}10}$ modes are the fundamental TM modes in the azimuthally modulated cavities that support them.

Additionally, we note that if the ratio of the transverse multipolar component to the monopolar component is zero, $\tilde{g}_{m_1}/\tilde{g}_0 = 0$, the form of $E_z$ in Eq. \ref{eq:MonoSingleMultipoleField} is equivalent to that of a TM$_{0n0}$ mode. Such modes are supported by a cavity with a circular cross-section, as discussed in Sec. \ref{sec:IsolatingSingle}. The cross-section of the cavity that supports a TM$_{\{0,m_1\}\eta 0}$ mode for a small ratio of transverse multipole to monopole, $\tilde{g}_{m_1}/\tilde{g}_0 << 1$, will therefore be perturbed only slightly from circular. Taking the other extreme whereby the transverse multipole to monopole ratio tends to infinity, $\tilde{g}_{m_1}/\tilde{g}_0 \rightarrow \infty$, the cavity cross-sections do not tend back to circular but instead tend to unique, non-circular shapes. This asymmetry arises because the TM$_{m_1n0}$ modes are degenerate whereas the TM$_{\{0,m_1\}\eta 0}$ modes are non-degenerate for all ratios of $\tilde{g}_{m_1}/\tilde{g}_0$.

Finally, we consider the azimuthally modulated cross-sections that support general TM$_{\{0,m_1\}\eta 0}$ modes for any value of $\phi_{m_1}$. Introducing a $\phi_{m_1}\neq0$ term into Eq. \ref{eq:MonoSingleMultipoleField} and Eq. \ref{eq:MonoSingleMultipole}, we find the set of general cavity cross-sections are identical to the subset of cavity cross-sections with $\phi_{m_1} = 0$, except for a physical rotation about the $z$-axis of $\phi_{m_1}/m_1$. 

\subsubsection{Example: Modes with a Monopolar and a Sextupolar Component} \label{sec:MonoSext}

It turns out that for certain combinations of $m_1$ and $\eta$, there is a limit on the ratio of transverse multipole to monopole beyond which the cross-section of the azimuthally modulated RF cavity becomes discontinuous. This is best illustrated through an example and here we consider 3 GHz TM$_{\{0,3\}\eta 0}$ modes with $\phi_3 = 0$. Inserting this, $m_1=3$ and $k  = 2\pi\times 3$ GHz$/c$ into Eq. \ref{eq:MonoSingleMultipole} gives the longitudinal electric field profile of such a mode:
\begin{equation}\label{eq:MonoSingleSextField}
    E_z(r,\theta,z) = J_0\left(k r\right)\tilde{g}_0 + J_3\left(k r\right) \tilde{g}_3\cos{3\theta},
\end{equation}
and the corresponding cross-section can be determined from Eq. \ref{eq:MonoSingleMultipole} which we rearrange to:
\begin{equation}\label{eq:MonoSextMultipole}
    J_0\left(k r_0^{(\eta)}(\theta)\right) = - \frac{\tilde{g}_3}{\tilde{g}_0}J_3\left(k r_0^{(\eta)}(\theta)\right)\cos{3\theta}.
\end{equation}

\begin{figure}
\subfloat[The cavity shapes, $r_0^{(\eta)}(\theta)$, that support the $\eta$ = 1, 2, and 3 TM$_{\{0,3\}\eta0}$ modes as black, blue, and cyan lines respectively.\label{fig:MonoSext0_95Shape}]{\includegraphics[width=0.475\textwidth]{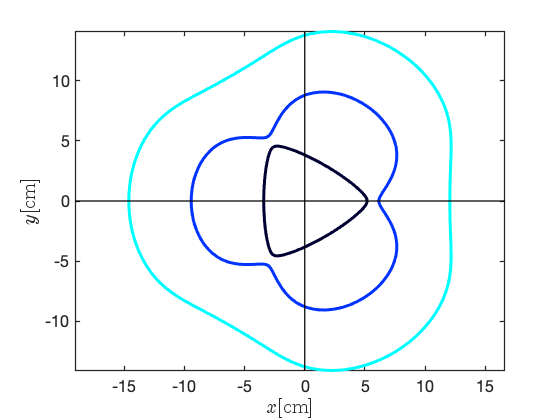}}\quad
\subfloat[$J_0(k r)$ and $-0.95J_3(k r)\cos{3\theta}$  plotted against $r$ for $\theta = 0, \pi/3$. The $\eta$ = 1, 2, and 3 solutions where the two functions intersect for all $\theta$ are plotted as black, blue, and cyan lines.\label{fig:MonoSext0_95Bess}]{\includegraphics[width=0.475\textwidth]{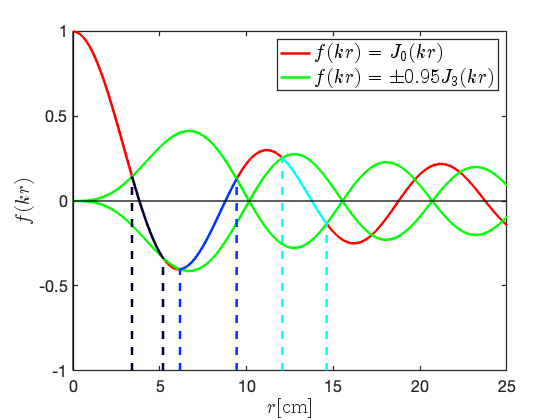}}
\caption{3 GHz TM$_{\{0,3\}\eta 0}$ modes with $\tilde{g}_3/\tilde{g}_0 = 0.95$ and $\phi_3 = 0$.}\label{fig:MonoSext0_95}
\end{figure}

We firstly solve Eq. \ref{eq:MonoSextMultipole} numerically for a sextupole to monopole ratio of $\tilde{g}_3/\tilde{g}_0$ = 0.95. The cross-sections of the RF cavities that support the $\eta=1$, 2, and 3  modes are plotted in Fig. \ref{fig:MonoSext0_95Shape}. They have $3$-fold rotational symmetry, as expected.

To help understand the properties of these azimuthally modulated cavities, we  analyse Eq. \ref{eq:MonoSextMultipole} graphically. Generalising the monopole term to $J_0(k r)$ and sextupole term to $-0.95J_3(k r)\cos{3\theta}$ then, if we plot these two functions against $r$ for a given $\theta=\theta_0$, Eq. \ref{eq:MonoSextMultipole} states the $r^{(\eta)}(\theta_0)$ solutions are the points at which the two functions intersect. Figure \ref{fig:MonoSext0_95Bess} plots $J_0(kr)$ in red and the two functions $\pm0.95J_3(k r)$ in green, the latter corresponding to the minimum and maximum values of $\cos{3\theta_0} = \mp1$. The $\eta$ = 1, 2, and 3 points of intersection of the red and green lines are indicated respectively by the black, blue, and cyan vertical dashed lines. 

Considering the intersections at other $\theta$ values, we note that the monopole term has no $\theta$ dependence and thus intersections will always lie on the red, monopolar line for all $\theta$. In contrast, the sextupolar term contains a cosine term that acts as a $\theta$-dependent amplifying term on $-0.95J_3(k r)$. As Fig. \ref{fig:MonoSext0_95Bess} plots $\pm 0.95J_3(k r)$ which correspond to the maximum and minimum amplification, intersections at all other $\theta$ must occur between these maximum and minimum intersections. The solid black, blue, and cyan lines on Fig. \ref{fig:MonoSext0_95Bess}  show the $\eta$ = 1, 2, and 3  intersections for all $\theta$ and it is clear that each $r^{(\eta)}(\theta)$ solution is bounded between its corresponding vertical dashed lines.

This graphical analysis of Eq. \ref{eq:MonoSextMultipole} in Fig. \ref{fig:MonoSext0_95Bess} explains why cavity shapes in Fig. \ref{fig:MonoSext0_95Shape} are physically larger for larger $\eta$. Additionally, Fig. \ref{fig:MonoSext0_95Bess} shows $j_{31} > j_{02}$, that is the 3$^{\textrm{rd}}$ order Bessel function first intersects the $r$-axis after the 0$^{\textrm{th}}$ order Bessel has intersected it twice. As a result, both the $\eta=1$  and $\eta=2$ cross-sections exist in the interval $r = [0,j_{31}]/k  \approx [0, 10]$ cm, and this explains why $r_0^{(1)}(0)$ is a maximum whereas $r_0^{(2)}(0)$ is a minimum in Fig. \ref{fig:MonoSext0_95Shape}. This is also why we could not equate $\eta$ with $x$ in Eq. \ref{eq:MonoSingleFreq} as there may be more than one solution in an interval $r = [j_{m_1(x-1)},j_{m_1x}]/k$.

\begin{figure}
\subfloat[The cavity shapes, $r_0^{(\eta)}(\theta)$, that support the $\eta$ = 1, 2, and 3 TM$_{\{0,3\}\eta0}$ modes as black, blue, and cyan lines respectively.\label{fig:MonoSext1_2Shape}]{\includegraphics[width=0.475\textwidth]{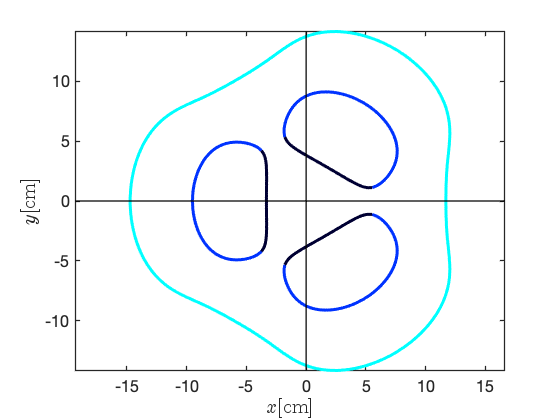}}\quad
\subfloat[$J_0(k r)$ and $-1.2J_3(k r)\cos{3\theta}$  plotted against $r$ for $\theta = 0, \pi/3$. The $\eta$ = 1, 2, and 3 solutions where the two functions intersect for all $\theta$ are plotted as black, blue, and cyan lines.\label{fig:MonoSext1_2Bess}]{\includegraphics[width=0.475\textwidth]{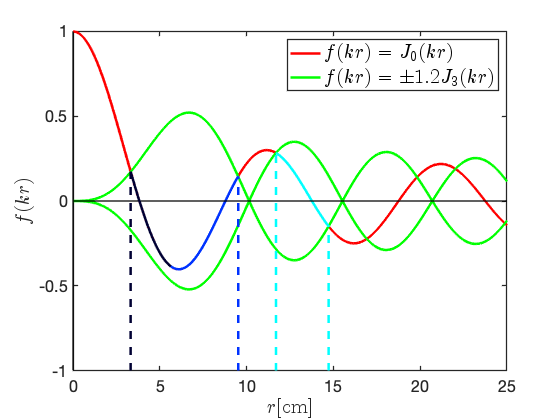}}
\caption{3 GHz TM$_{\{0,3\}\eta 0}$ modes with $\tilde{g}_3/\tilde{g}_0 = 1.2$ and $\phi_3 = 0$.}\label{fig:MonoSext1_2}
\end{figure}

There is a further consequence from having multiple solutions in an interval $r = [j_{m_1(x-1)},j_{m_1x}]/k$: if we consider increasing the value of $\tilde{g}_3/\tilde{g}_0$ beyond 0.95 then the $\eta$ = 1 and $\eta$ = 2 solutions at the angles $\cos{3\theta_0} = 1$  will further converge  until, at a critical ratio of $\tilde{g}_3/\tilde{g}_0$, $r_0^{(1)}(\theta_0)$ = $r_0^{(2)}(\theta_0)$. To see this graphically, consider the green line representing $-0.95J_3(k r)$ in Fig. \ref{fig:MonoSext0_95Bess}: if $\tilde{g}_3/\tilde{g}_0$ was increased from 0.95, the $\eta = 1$ (black) and $\eta = 2$ (blue) intersection points at $r\approx5.5$ cm will move closer together. The critical value at which the $r_0^{(1)}(\theta_0)$ and $r_0^{(2)}(\theta_0)$ converge is numerically calculated as  $(\tilde{g}_3/\tilde{g}_0)_\textrm{crit}$ = 0.98. 

The implication of trying to solve Eq. \ref{eq:MonoSextMultipole} with $\tilde{g}_3/\tilde{g}_0 = 1.2 > (\tilde{g}_3/\tilde{g}_0)_\textrm{crit}$ is shown in Fig. \ref{fig:MonoSext1_2}: for a range of $\theta$ values around $\cos{3\theta}=0$, real $\eta=1$ and $\eta=2$ solutions do not exist because there is no intersection between the Bessel terms, and therefore solution, at $r \approx 5.5$ cm. As a result, the $r_0^{(1)}(\theta)$ and $r_0^{(2)}(\theta)$ solutions are discontinuous. 

This can be contrasted to the  continuous $r_0^{(3)}(\theta)$ solution in Fig. \ref{fig:MonoSext1_2Bess} whereby the  two terms continuously intersect for all $\theta$. In fact, because there is only one root to the $0^\textrm{th}$ order Bessel function in the interval $[j_{31},j_{32}]$, the $\eta = 3$ solution is continuous for all values of  $\tilde{g}_3/\tilde{g}_0$. We can also see that the $\eta = 3$ solution is bounded within the interval $[j_{31},j_{32}]/k$. Specifically as $\tilde{g}_3/\tilde{g}_0 \rightarrow 0$, $r_0^{(3)}(\theta)\rightarrow j_{03}/k$, and as $\tilde{g}_3/\tilde{g}_0 \rightarrow \infty$, $r_0^{(3)}(\theta)\rightarrow [j_{31}/k$ for $\cos{3\theta}>0$; $j_{32}/k$ for $\cos{3\theta}<0$].

The mathematical solutions to Eq. \ref{eq:EzToSolve} can inform the practical and realistic design of cavities that support TM$_{\{M\}\eta 0}$-like modes for useful application in particle accelerators. RF cavities used in particle accelerators must have an enclosed cross-section and may have constraints on their maximum physical size and frequency of operation. Therefore, although the $r_0^{(3)}(\theta)$ structure is continuous and enclosed for the multipolar ratio  $\tilde{g}_3/\tilde{g}_0=1.2$, it is a physically larger cavity than the discontinuous $r_0^{(1)}(\theta)$ and $r_0^{(2)}(\theta)$ shapes. As a potential compromise between creating a mode with the desired  $\tilde{g}_3/\tilde{g}_0=1.2$ multipolar ratio whilst minimising the physical size of the cavity that supports it, we can consider designing an RF cavity where the cross-section jumps to the $\eta=3$ solution at the angles where real $\eta=1$ and $\eta=2$ solutions do not exist. We define such cavities that mix modes of differing $\eta$ as $\textit{hybrid}$ cavities and these are further discussed and analysed in Sec. \ref{sec:CSTSimulations}.

\subsubsection{Conditional Modes} \label{sec:SingleMonoConditional}

For the case of the TM$_{\{0,3\}\eta 0}$ modes, we showed that  the $\eta=1$ and $\eta=2$ solutions are discontinuous if $\tilde{g}_3/\tilde{g}_0$ exceeds a critical value. This arises  because $j_{31}>j_{02}$ and so there are two roots to the $0^\textrm{th}$ order Bessel function in the interval $[0,j_{31}]$. In contrast, the $\eta=3$ solution is continuous for all values of $\tilde{g}_3/\tilde{g}_0$ because there is only one root to the $0^\textrm{th}$ order Bessel function in the interval $[j_{31},j_{32}]$. Extending this, the azimuthally modulated cavity designed to support a TM$_{\{0,m_1\}\eta 0}$ mode will have a critical limit on $\tilde{g}_{m_1}/\tilde{g}_0$ beyond which the solution is discontinuous if in the $[j_{m_1(x-1)}, j_{m_1x}]$ interval in which the $r_0^{(\eta)}(\theta)$ solution lies, there is more than one root to the $0^{\textrm{th}}$ order Bessel function.  

To  generalise this result, we define any TM$_{\{M\}\eta 0}$ mode of given multipolar content $\{M\}$ and radial order $\eta$ as \textit{conditional} if any corresponding $r_0^{(\eta)}(\theta)$ solution is discontinuous for any possible combination of magnitudes of the multipolar coefficients. A TM$_{\{M\}\eta0}$ mode is therefore \textit{unconditional} if it has a continuous cross-section for all possible ratios of the multipolar coefficients.

\subsection{Modes with Two Different Transverse Multipolar Components} \label{sec:TwoMultipoles}

Having determined the properties of azimuthally modulated RF cavities that support the TM$_{\{0,m_1\}\eta 0}$ modes in the previous subsection, we now examine the azimuthally modulated RF cavities that support the TM$_{\{m_1,m_2\}\eta 0}$ modes composed of two transverse multipoles. 

Using Eq. \ref{eq:EzForm}, the TM$_{\{m_1,m_2\}\eta 0}$ modes composed of two different, transverse multipoles with $m_2>m_1>0$ have a longitudinal electric field of the form:
\begin{multline}\label{eq:TwoMultipoleField}
    E_z(r,\theta,z) = J_{m_1}\left(k r\right)\tilde{g}_{m_1}\cos{(m_1\theta-\phi_{m_1})} 
     + \\J_{m_2}\left(k r\right) \tilde{g}_{m_2}\cos{(m_2\theta-\phi_{m_2})}.
\end{multline}
Without loss of generality, the corresponding set of azimuthally modulated cavity shapes can be determined from Eq. \ref{eq:EzToSolve} with $\phi_{m_1} = 0$:
\begin{multline}\label{eq:TwoMultipole}
    J_{m_1}\left(k r_0^{(\eta)}(\theta)\right)\tilde{g}_{m_1}\cos{m_1\theta} +\\ 
    J_{m_2}\left(k r_0^{(\eta)}(\theta)\right)\tilde{g}_{m_2}\cos{(m_2\theta-\phi_{m_2})}= 0.
\end{multline}

The cross-sections that solve Eq. \ref{eq:TwoMultipole} and support TM$_{\{m_1,m_2\}\eta 0}$ modes require further consideration compared to those that support TM$_{\{0,m_1\}\eta 0}$ modes for four interlinked reasons:

\begin{itemize}
    \item Solutions do not necessarily have an $m_1$-fold symmetry but instead a rotational symmetry that varies depending upon the values of $m_1$, $m_2$ and $\phi_{m_2}$. As a result, $\eta$ has a more specific definition for TM$_{\{m_1,m_2\}\eta 0}$ modes.
    \item A practical RF cavity must have $r_0^{(\eta)}(\theta) >0$, however, $r_0^{(\eta)}(\theta) = 0$ is a valid solution to Eq. \ref{eq:TwoMultipole} for all $m_1$ and $m_2$. We therefore define a TM$_{\{m_1,m_2\}\eta 0}$ mode as \textit{forbidden} if $r_0^{(\eta)}(\theta) = 0$ at any angle $\theta$.
    \item Solutions are not necessarily bounded within the interval $r = [j_{m_2(x-1)}$, $j_{m_2x}]/k$ due to both terms in Eq. \ref{eq:TwoMultipole} having differing orders of cosine dependence. 
    \item A practical RF cavity must be closed with $r^{(\eta)}(0)=r^{(\eta)}(2\pi)$, however, solutions of particular $m_1$, $m_2$ and $\phi_{m_2}$ do not satisfy this condition. We define such TM$_{\{m_1,m_2\}\eta0}$ modes as \textit{spiral}.
\end{itemize}

Appendix \ref{sec:AppForbiddenSpiral} derives why certain modes are forbidden or spiral. Section \ref{sec:IsolatingQuadOct} presents an example of a forbidden mode and Sec. \ref{sec:Spiral} an example of a spiral mode. 

TM$_{\{m_1,m_2\}\eta 0}$ modes can also be conditional whereby the cross-section is discontinuous and a hybrid solution needs be considered if the ratio of the transverse multipolar coefficient of order $m_1$ to the transverse multipolar coefficient of order $m_2$ (or vice versa) exceeds a critical value. A TM$_{\{m_1,m_2\}\eta 0}$ mode will be conditional if, for any $\theta=\theta_0$, there is more than one root to the $m_1^\textrm{th}$ order Bessel function within the $[j_{m_2(x-1)}, j_{m_2x}]/k$ interval that the $r_0^{(\eta)}(\theta_0)$ solution exists in.

\subsubsection{Example: Modes with a Quadrupolar and an Octupolar Component} \label{sec:IsolatingQuadOct}
\begin{figure}
\subfloat[The cavity shapes, $r_0^{(\eta)}(\theta)$, that support the $\eta$ = 1, 2, and 3 TM$_{\{2,4\}\eta0}$ modes as black, blue, and cyan lines respectively.\label{fig:QuadOctShape}]{\includegraphics[width=0.475\textwidth]{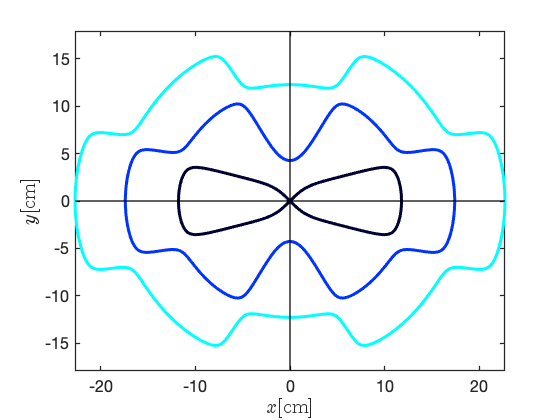}}\quad
\subfloat[$\pm J_2(k r)$ and $\pm5J_4(k r)$  plotted against $r$. The $\eta$ = 1, 2, and 3 solutions where the  $J_2(k r)\cos2\theta$ and $-5J_4(k r)\cos4\theta$ functions intersect for all $\theta$ are plotted as black, blue, and cyan lines.\label{fig:QuadOctBessels}]{\includegraphics[width=0.475\textwidth]{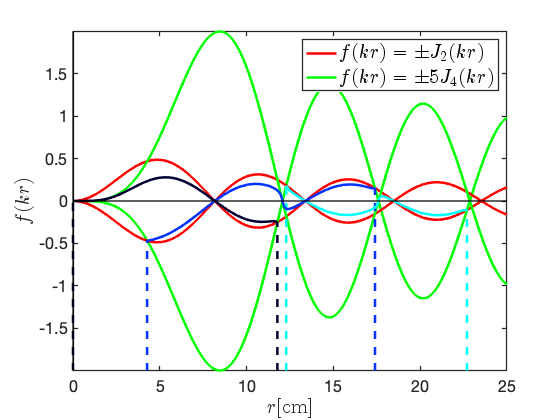}}
\caption{3 GHz TM$_{\{2,4\}\eta 0}$ modes with $\tilde{g}_4/\tilde{g}_2 = 5$, $\phi_4=0$.}\label{fig:QuadOct}
\end{figure}
To illustrate a forbidden mode, consider the $\eta$ = 1, 2, and 3 cross-sections that support 3 GHz TM$_{\{2,4\}\eta0}$ modes with $\tilde{g}_4/\tilde{g}_2$ = 5 and $\phi_{m_2}=0$.  Numerically solving Eq. \ref{eq:TwoMultipole} with these parameters returns the cross-sections plotted in Fig. \ref{fig:QuadOctShape}. Figure \ref{fig:QuadOctBessels} plots the two Bessel terms for maximum and minimum amplitude and the $\eta$ = 1, 2, and 3 solutions. It shows that each solution crosses one interval and so the $\eta=1$ solution forbidden. Additionally,  because there is only a single root to the 2$^\textrm{nd}$ order Bessel function within each $[j_{4(x-1)},j_{4x}]/k$ interval, all TM$_{\{2,4\}\eta 0}$ modes are unconditional and each solution bounded between $[j_{4(\eta-2)},j_{4\eta}]/k$.

\subsubsection{Example: Modes with a Dipolar and Sextupolar Component} \label{sec:Spiral}

\begin{figure}
    \centering
       \includegraphics[width=0.5\textwidth]{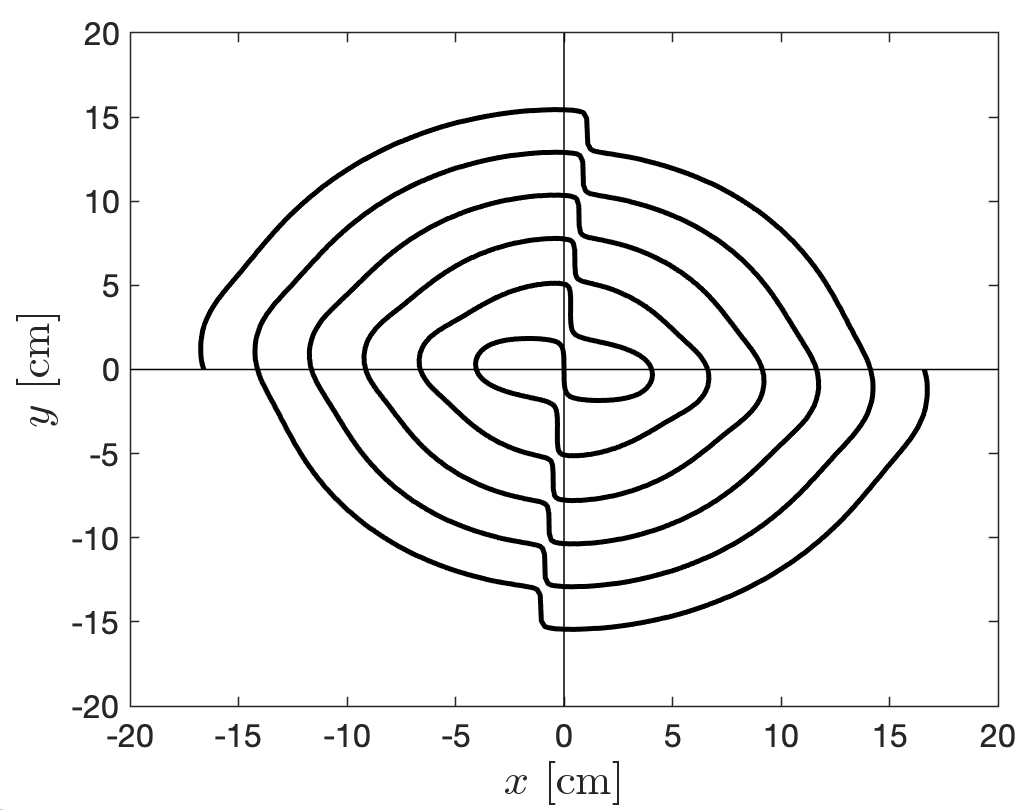}
        \caption{The $r^{(\eta)}(\theta)$ solutions that support the 3 GHz TM$_{\{1,3\}\eta 0}$ modes with $\tilde{g}_3/\tilde{g}_1 = 1$ and $\phi_3 = 0.3$ is plotted in black in the range $r \approx [0, 15]$ cm.}  \label{fig:Spiral}
\end{figure}

Figure \ref{fig:Spiral} shows an example of a spiral mode by plotting the $r_0^{(\eta)}(\theta)$ solutions in the range $r \approx [0, 15]$ cm that support 3 GHz TM$_{\{1,3\}\eta0}$ modes with $\tilde{g}_3/\tilde{g}_1$ = 1 and $\phi_3 = 0.3$. The solutions form a single, continuous, spiralling line which differs from the discrete, closed solutions formed by unconditional solutions of distinct $\eta$ in previous examples. It is evident that $r_0^{(\eta)}(0)\neq r_0^{(\eta)}(2\pi)$ and therefore this is a spiral mode. 

\subsection{Modes with Three Or More Different Multipolar Components}

Equation \ref{eq:EzToSolve} can also be numerically solved to determine the azimuthally modulated cross-sections that support TM$_{\{M\}\eta0}$ modes with three or more multipolar components. General mathematical analysis of the properties of these modes and the corresponding cross-sections that support them is challenging. The concepts of certain modes being conditional, forbidden or spiral, however,  generalises to TM$_{\{M\}\eta0}$ modes with any number multipolar content and provides rationale for understanding the numerical solutions to Eq. \ref{eq:EzToSolve} as well as helping to inform the scope and limitation of a general azimuthally modulated RF cavity design.

\section{CST Simulations} \label{sec:CSTSimulations}

\begin{figure*}
\subfloat[][TM$_{310}$.  \phantom{g}\label{subfig:CSTF0}]{\includegraphics[width=0.17\textwidth]{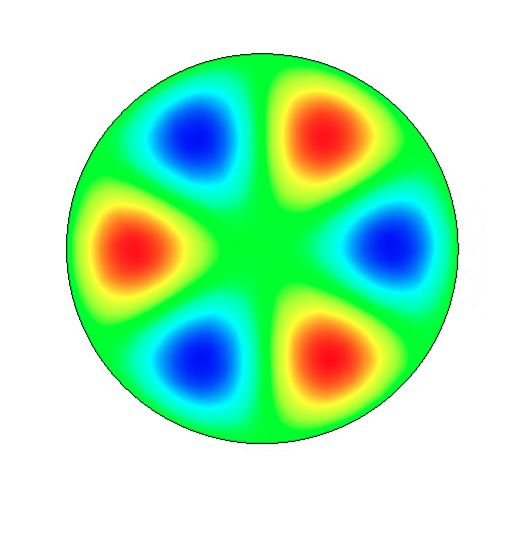}}
\subfloat[][TM$_{\{0,3\}10}$,  $\tilde{g}_3/\tilde{g}_0 = 0.95.$\label{subfig:CSTF1}]{\includegraphics[width=0.17\textwidth]{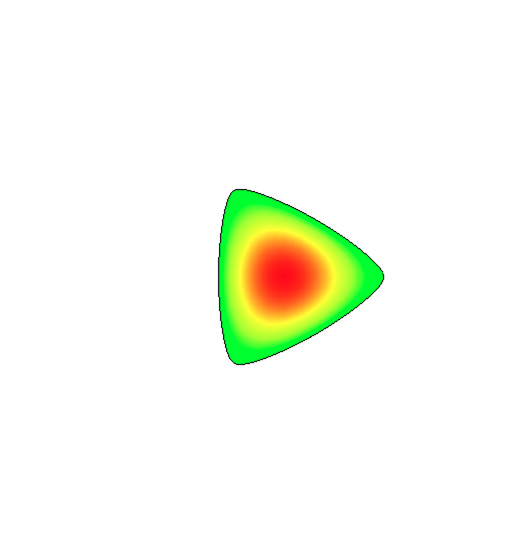}}
\subfloat[][TM$_{\{0,3\}20}$,  $\tilde{g}_3/\tilde{g}_0 = 0.95.$\label{subfig:CSTF2}]{\includegraphics[width=0.17\textwidth]{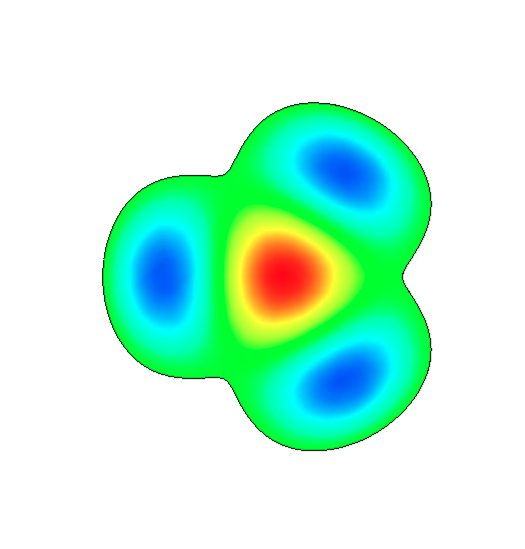}}
\subfloat[][TM$_{\{0,3\}30}$,  $\tilde{g}_3/\tilde{g}_0 = 0.95.$\label{subfig:CSTF3}]{\includegraphics[width=0.17\textwidth]{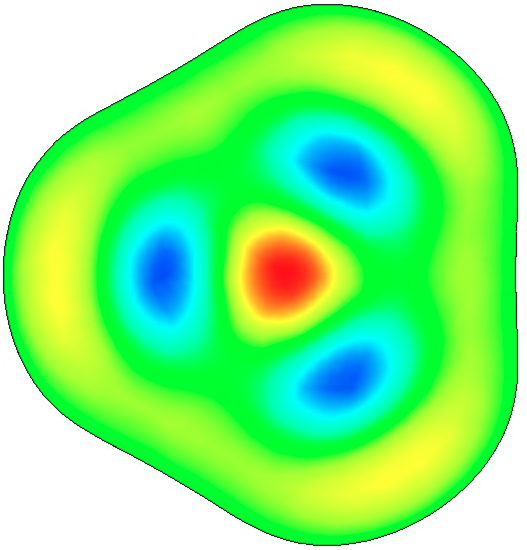}}
\subfloat[][Hybrid TM,  $\tilde{g}_3/\tilde{g}_0 = 1.2.$\label{subfig:CSTF4}]{\includegraphics[width=0.17\textwidth]{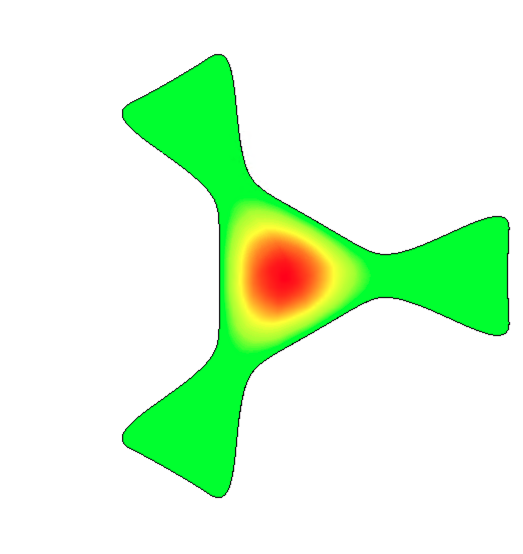}}
\caption{$E_z$ in cavities operating in 3 GHz TM$_{310}$, TM$_{\{0,3\}\eta 0}$ and hybrid TM$_{\{0,3\}1 0}$-TM$_{\{0,3\}3 0}$ modes. Red is a maximum in the field, blue a minimum and green is zero. The cavities are to scale: the radius of the circular pillbox is 10.15 cm.}\label{fig:CSTFields}
\end{figure*}

\begin{figure*}
\subfloat[][TM$_{310}$.  \phantom{g}\label{subfig:CSTD0}]{\includegraphics[width=0.175\textwidth]{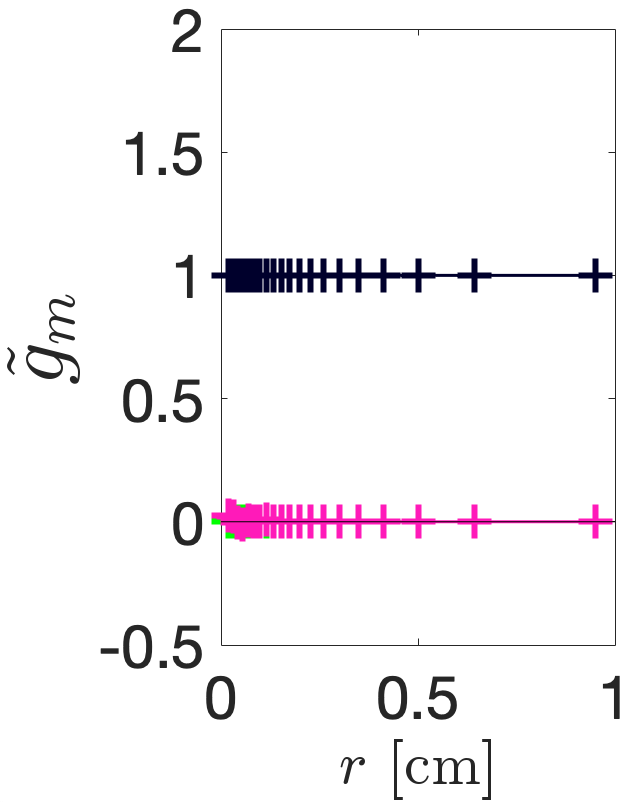}}
\subfloat[][TM$_{\{0,3\}10}$,  $\tilde{g}_3/\tilde{g}_0 = 0.95.$\label{subfig:CSTD1}]{\includegraphics[width=0.175\textwidth]{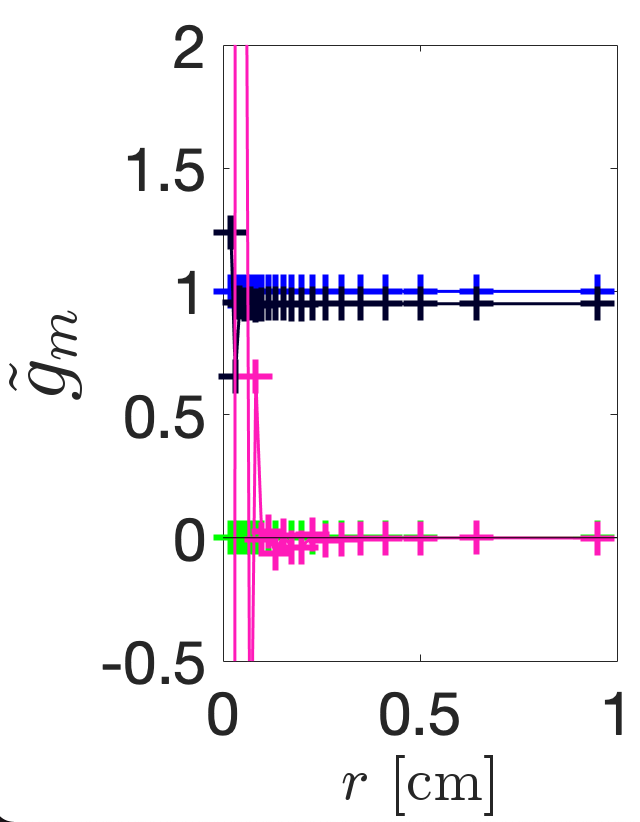}}
\subfloat[][TM$_{\{0,3\}20}$,  $\tilde{g}_3/\tilde{g}_0 = 0.95.$\label{subfig:CSTD2}]{\includegraphics[width=0.175\textwidth]{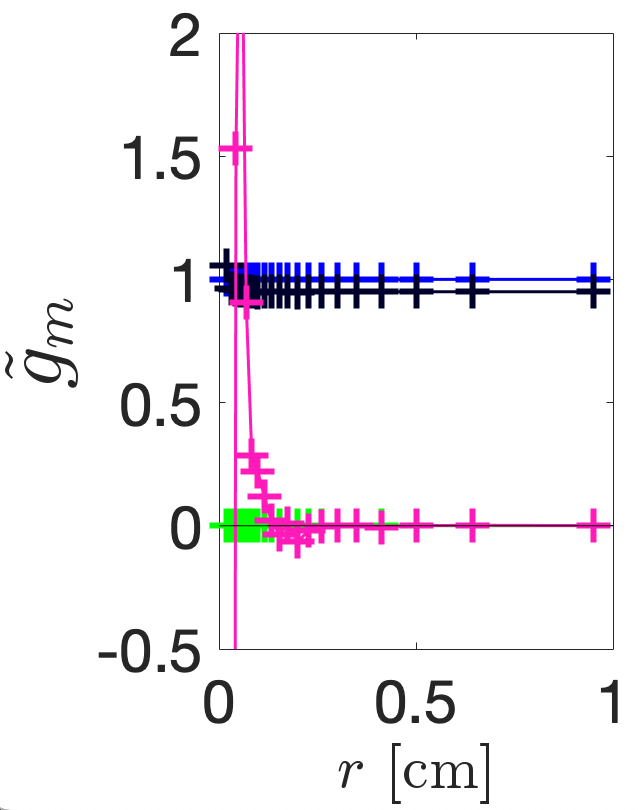}}
\subfloat[][TM$_{\{0,3\}30}$,  $\tilde{g}_3/\tilde{g}_0 = 0.95.$\label{subfig:CSTD3}]{\includegraphics[width=0.175\textwidth]{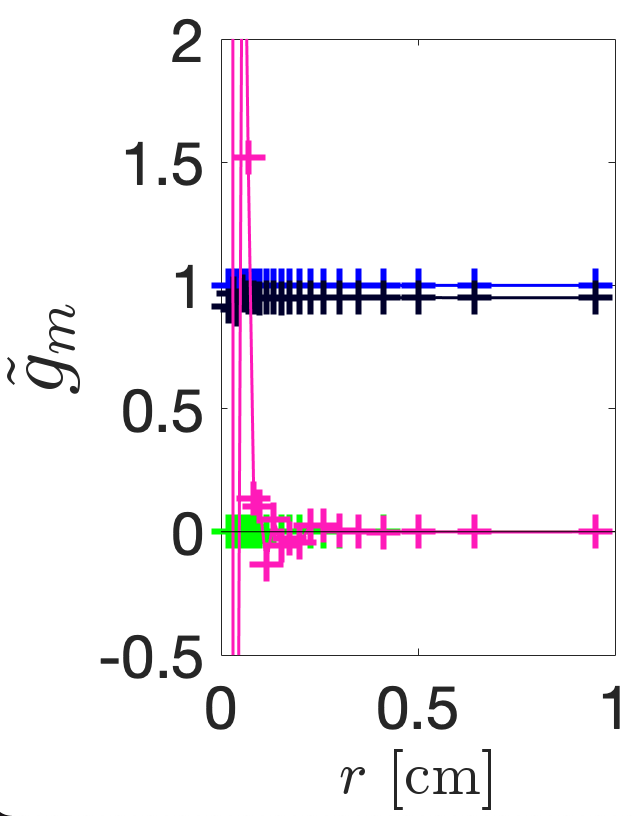}}
\subfloat[][TM$_{\{0,3\}10}$,  $\tilde{g}_3/\tilde{g}_0 = 1.2.$\label{subfig:CSTD4}]{\includegraphics[width=0.175\textwidth]{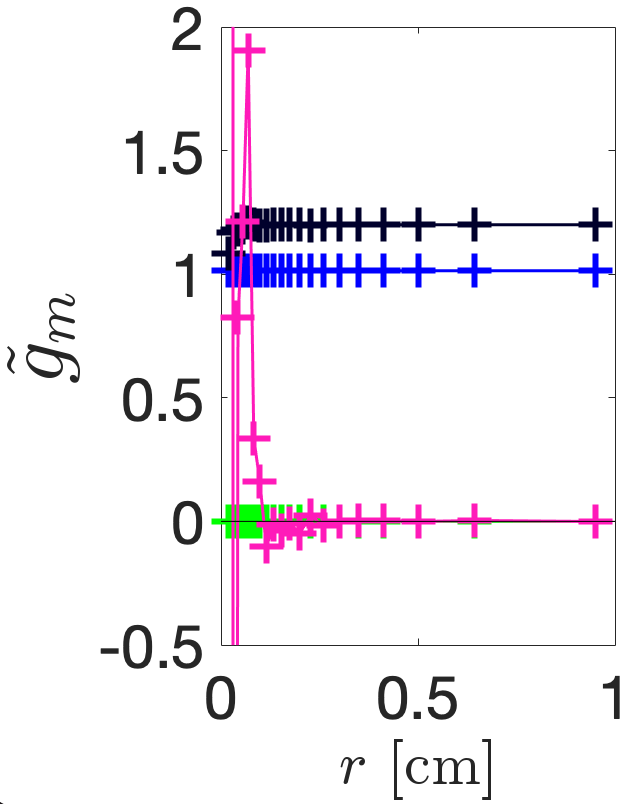}}
\subfloat{\includegraphics[width=0.06\textwidth]{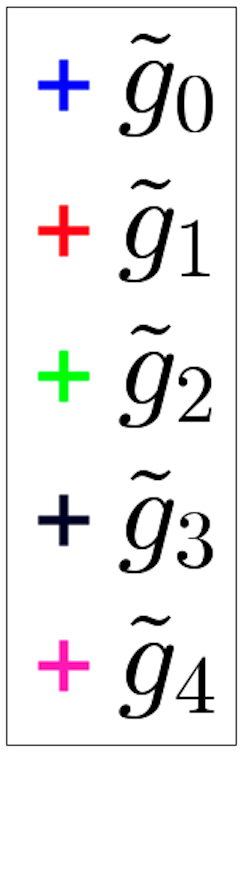}}
\caption{The calculated multipolar coefficient, $\tilde{g}_m$, is plotted against the distance from the centre, $r$, with colour denoting the order, $m$, for cavities operating in the TM$_{310}$ and four different TM$_{\{0,3\}\eta0}$ modes.}\label{fig:CSTDecomp}
\end{figure*}

\begin{figure*}[tbh!]
\subfloat[][TM$_{310}$.  \phantom{g}\label{subfig:CSTO0}]{\includegraphics[width=0.19\textwidth]{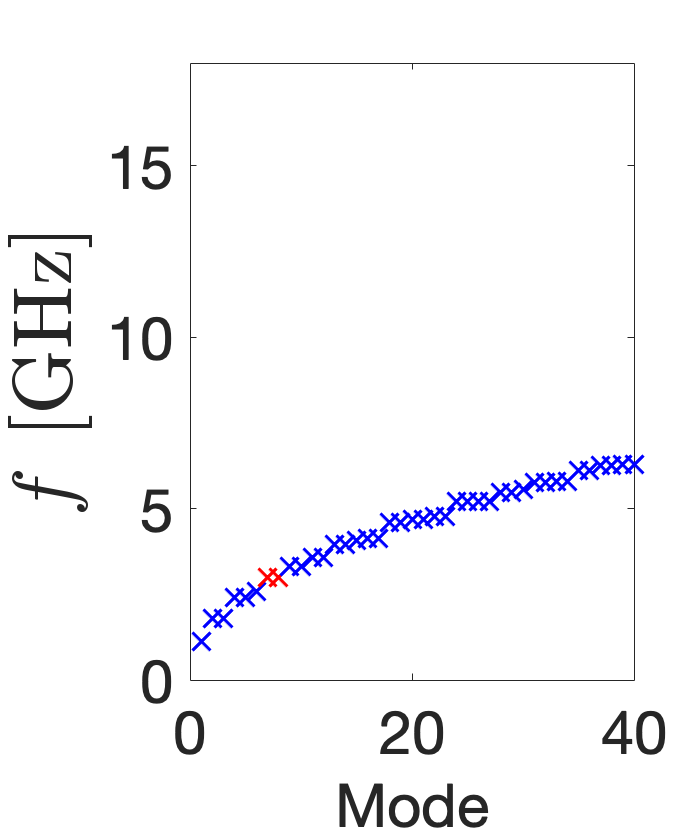}}
\subfloat[][TM$_{\{0,3\}10}$,  $\tilde{g}_3/\tilde{g}_0 = 0.95.$\label{subfig:CSTO1}]{\includegraphics[width=0.19\textwidth]{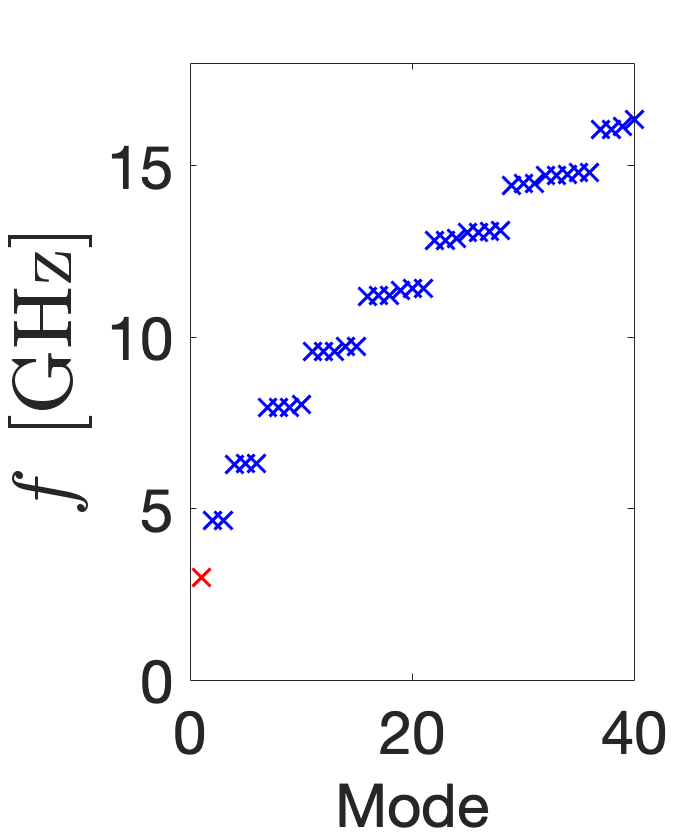}}
\subfloat[][TM$_{\{0,3\}20}$,  $\tilde{g}_3/\tilde{g}_0 = 0.95.$\label{subfig:CSTO2}]{\includegraphics[width=0.19\textwidth]{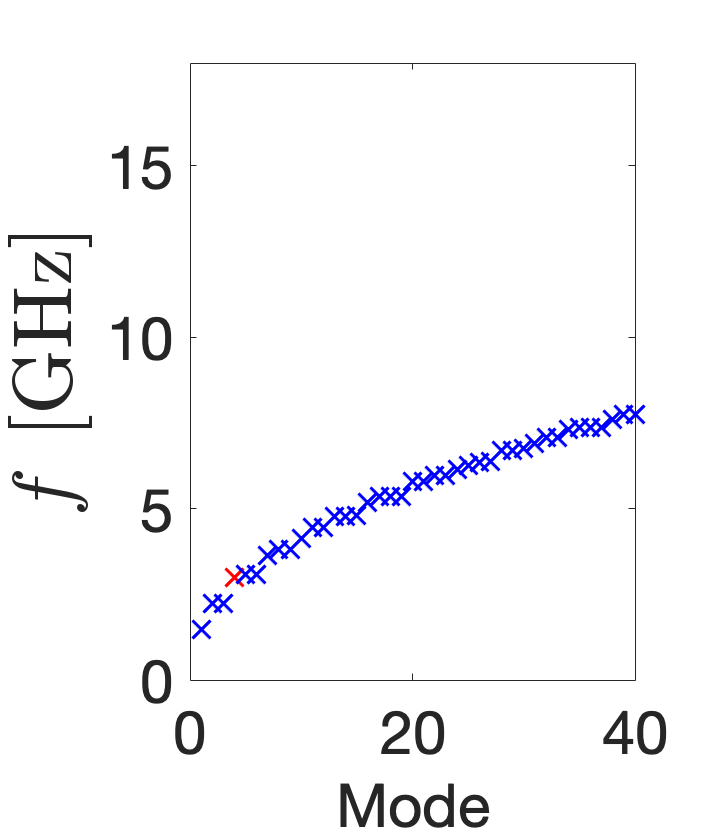}}
\subfloat[][TM$_{\{0,3\}30}$,  $\tilde{g}_3/\tilde{g}_0 = 0.95.$\label{subfig:CSTO3}]{\includegraphics[width=0.19\textwidth]{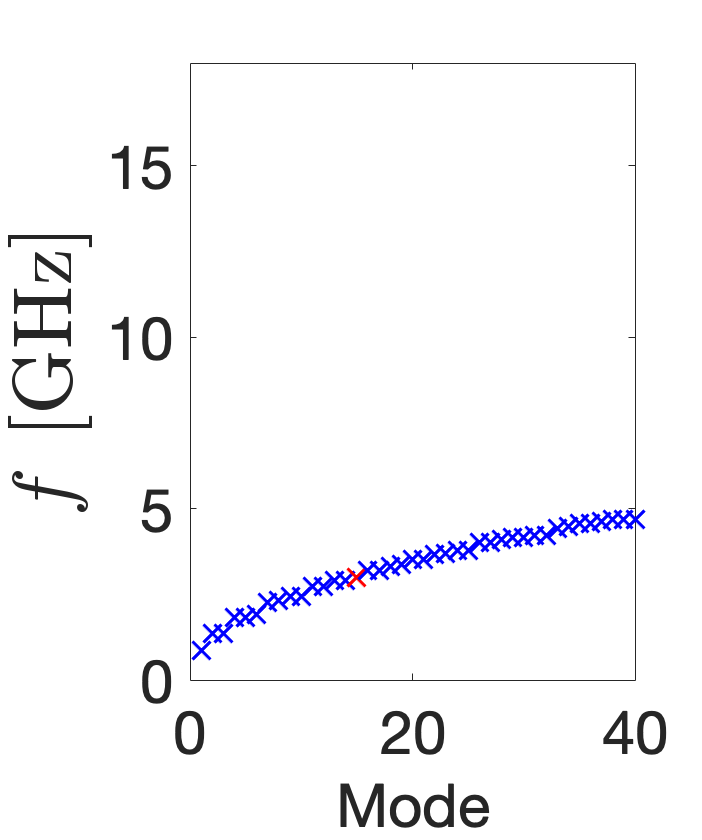}}
\subfloat[][TM$_{\{0,3\}10}$,  $\tilde{g}_3/\tilde{g}_0 = 1.2.$\label{subfig:CSTO4}]{\includegraphics[width=0.19\textwidth]{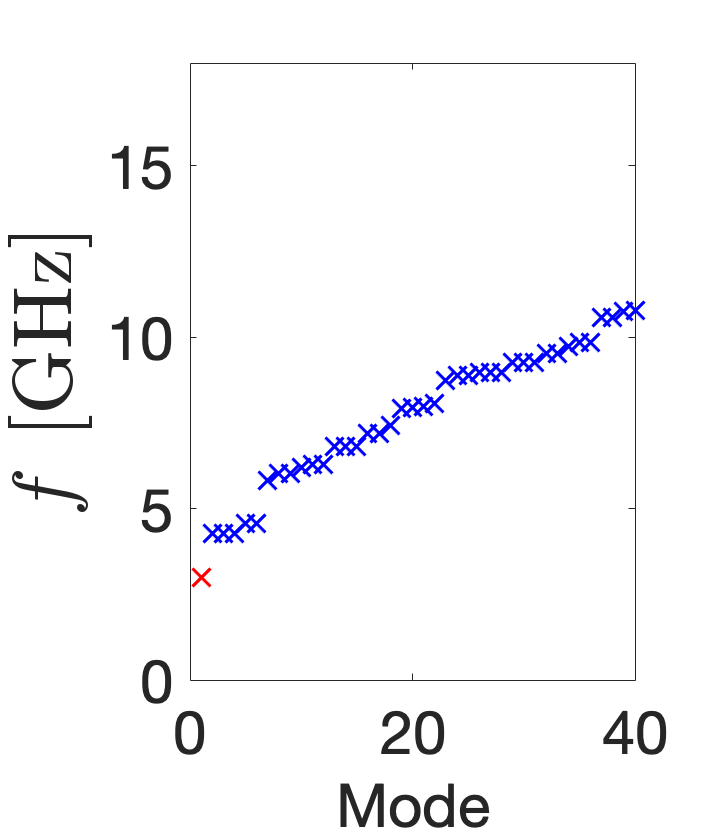}}
\caption{Frequency spectra of the 40 lowest frequency TM modes for the 5 different cavity shapes that support the mode denoted in the subcaption. The red crosses represent the 3 GHz mode of interest.}\label{fig:CSTFreqs}
\end{figure*}

Analysis of 3D electromagnetic simulations of azimuthally modulated RF cavities is presented here to confirm the validity of the theoretical work. Electric field distributions, Helmholtz decomposition results, and mode spectra are shown for: a pillbox cavity operating in a TM$_{310}$ mode, for azimuthally modulated cavities with continuous cross-sections that support the TM$_{\{0,3\}10}$, TM$_{\{0,3\}20}$, and TM$_{\{0,3\}30}$ modes with $\tilde{g}_3/\tilde{g}_0 = 0.95$, and for an azimuthally modulated hybrid cavity that supports a hybrid TM$_{\{0,3\}10}$-TM$_{\{0,3\}30}$ mode with $\tilde{g}_3/\tilde{g}_0 = 1.2$. All modes are normal with $\phi_3 = 0 $ and each of the cavities were designed with length 0.5 mm and such that the mode of interest resonates at 3 GHz. 
 
To create the hybrid cavity, the cavity wall is extended from the $r_0^{(1)}(\theta)$ solution to the $r_0^{(3)}(\theta)$ solution at the angles where the $\eta=1$ solution is no longer real. Although this creates a closed cavity, it introduces additional transverse multipoles into the mode because the boundary condition $E_z(r,\theta,z)=0$ must also be satisfied where the cavity wall extends from the $\eta=1$ to the $\eta=3$ solution. The field is small in this region compared to the field at the centre, however, and so the additional multipoles are also small. Thus the hybrid mode is dominantly, but not solely, composed of monopolar and sextupolar components and the cavity will resonate close to, but not exactly at, the designed resonant frequency. 

Figure \ref{fig:CSTFields} shows the $E_z$ distribution of these different modes. The difference between the TM$_{310}$ mode and TM$_{\{0,3\}\eta0}$ modes is evident by comparing the subfigures: there is no longitudinal accelerating field at the centre of the circular pillbox cavity operating in the TM$_{310}$ mode (Fig. \ref{subfig:CSTF0}) whereas there is in the case of the TM$_{\{0,3\}\eta0}$ modes (Figs. \ref{subfig:CSTF1}-\ref{subfig:CSTF4}). The resonant frequency of each of the modes was calculated by CST as 3.000 GHz, except for the hybrid cavity which resonated at 3.004 GHz.

Figure \ref{fig:CSTDecomp} shows the multipolar content of each of the five fields calculated using the Helmholtz decomposition method. This method is detailed in Ref. \cite{LongTermDynamics} and involves exporting the longitudinal electric field along the surface of cylinders orientated along the cavity axis and then performing an FFT on the data to determine $\tilde{g}_m$. To determine any dependence of these coefficients with radial distance, $E_z$ is exported along 20 discretely formed cylinders of non-linearly spaced radii in the range [0.02, 0.95] cm. By normalising the results to $\tilde{g}_3$ in Fig. \ref{subfig:CSTD0}, it is clear that the TM$_{310}$ mode in the circular pillbox only contains a sextupole component that is constant with radius. By normalising the results to $\tilde{g}_0$ in Figs. \ref{subfig:CSTD1}-\ref{subfig:CSTD4} we find the calculated multipolar content to be as designed: for example, in Fig. \ref{subfig:CSTD1}, $\tilde{g}_0 = 1$ for all radii and $\tilde{g}_3 = 0.95$ for all radii beyond $r \approx 0.03$ cm. The reason for the sextupolar and octupolar components diverging randomly at small $r$ from the expected constant value is because the data must be divided by $r^m$ to determine $\tilde{g}_m$. Simulation errors arising from the finite accuracy of the mesh and numerical errors arising from the pseudo-cylinders being  composed only of 360 distinct points are therefore amplified for smaller radii and larger $m$. Accounting for this error, the multipoles are constant with radial distance for each cavity shape and thus simulation results agree with the radial Bessel dependence given by Eq. \ref{eq:EzForm}.

The azimuthally modulated RF cavity shapes, $r_0^{(\eta)}(\theta)$, support the infinite number of modes that satisfy their boundary conditions. Figure \ref{fig:CSTFreqs} plots the frequency spectra of the 40 lowest frequency TM modes of the five different cavity shapes. The red crosses denote the 3 GHz mode of interest and any orthogonal degenerate modes of the same frequency. Figure \ref{subfig:CSTO0} shows the TM$_{310}$ mode is not the fundamental but the 7$^\textrm{th}$ lowest mode and that it is degenerate with the skew mode, as expected. This can be contrasted with the TM$_{\{0,3\}10}$ modes in Fig. \ref{subfig:CSTO1} and Fig. \ref{subfig:CSTO4}: not only are these modes fundamental and non-degenerate, but they also have a larger frequency separation. Additionally, while the frequency spectra of the TM$_{\{0,3\}20}$ and TM$_{\{0,3\}30}$ modes is similar in distribution to that of the circular pillbox and the modes are not fundamental, they are evidently non-degenerate with no same-order mode. The frequency spectra of the cavities is important because other modes may require consideration in a particle accelerator if they are excited, directly by the RF power supply or indirectly by a passing beam inducing wakefields, and this significantly interferes with beam dynamics. Any interfering modes may require damping by incorporating lower- and higher-order mode couplers into the design or could necessitate  a complete redesign of the cavity.

\section{Experimental Test of an RF Cavity that Simultaneously  Accelerates Longitudinally and  Focuses Transversely} \label{sec:QuadBuild}

Here in Sec. \ref{sec:QuadBuild}, we show that beam pipes and a dual-port power coupler can be incorporated into an azimuthally modulated RF cavity design whilst maintaining desired multipolar ratios. We also test a prototype in order to verify that the bead pull method can be used to measure the field of an azimuthally modulated cavity and accurately determine the ratio of multipolar terms with minimal error.  


\subsection{Designing the Cavity}

\begin{figure}[h]
    \centering
       \includegraphics[width=0.5\textwidth]{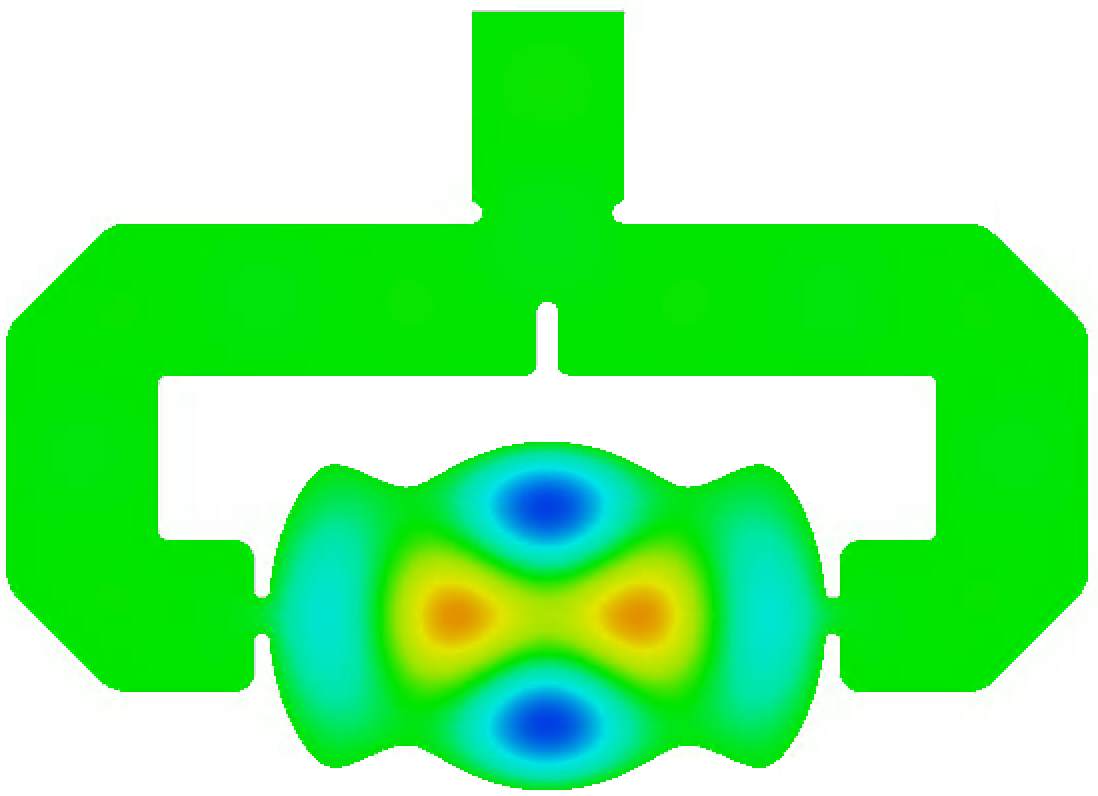}
        \caption{$E_z$ at the longitudinal centre of the designed RF cavity system, where blue is a minimum in the field, orange maximum, and green zero.}  \label{fig:QuadBuildCST}
\end{figure}

Figure \ref{fig:QuadBuildCST} shows the E-field distribution of a critically-coupled, single-cell, azimuthally modulated RF cavity system designed to support a 3 GHz TM$_{\{0,2\}20}$ mode with a quadrupole to monopole ratio of 5.041. Instead of using a single-port coupling system which would introduce a dipolar component into the field of the RF cavity, a dual-port coupling system was used in conjunction with a T-junction waveguide design to split the input power equally into the two waveguide arms. The dual-port coupler introduces an additional quadrupolar component into the mode. Helmholtz decompositions of simulations showed that by solving Eq. \ref{eq:EzToSolve} for the enclosed RF cavity shape that supports a 3 GHz TM$_{\{0,2\}20}$ mode with $\tilde{g}_2/\tilde{g}_0$ = 4.032 and $\phi_2 = 0$, and using this shape to create the prototype in Fig. \ref{fig:QuadBuildCST} which includes the dual port coupler and holes at the positions of the poles, the resultant field had a quadrupole to monopole ratio of $\tilde{g}_2/\tilde{g}_0$ = 5.041 and negligible contributions from higher order multipoles. The $\eta=2$ mode was chosen over the $\eta=1$ mode in order to give more measurement points to consider and compare.

\begin{figure}[h]
    \centering
       \includegraphics[width=0.48\textwidth]{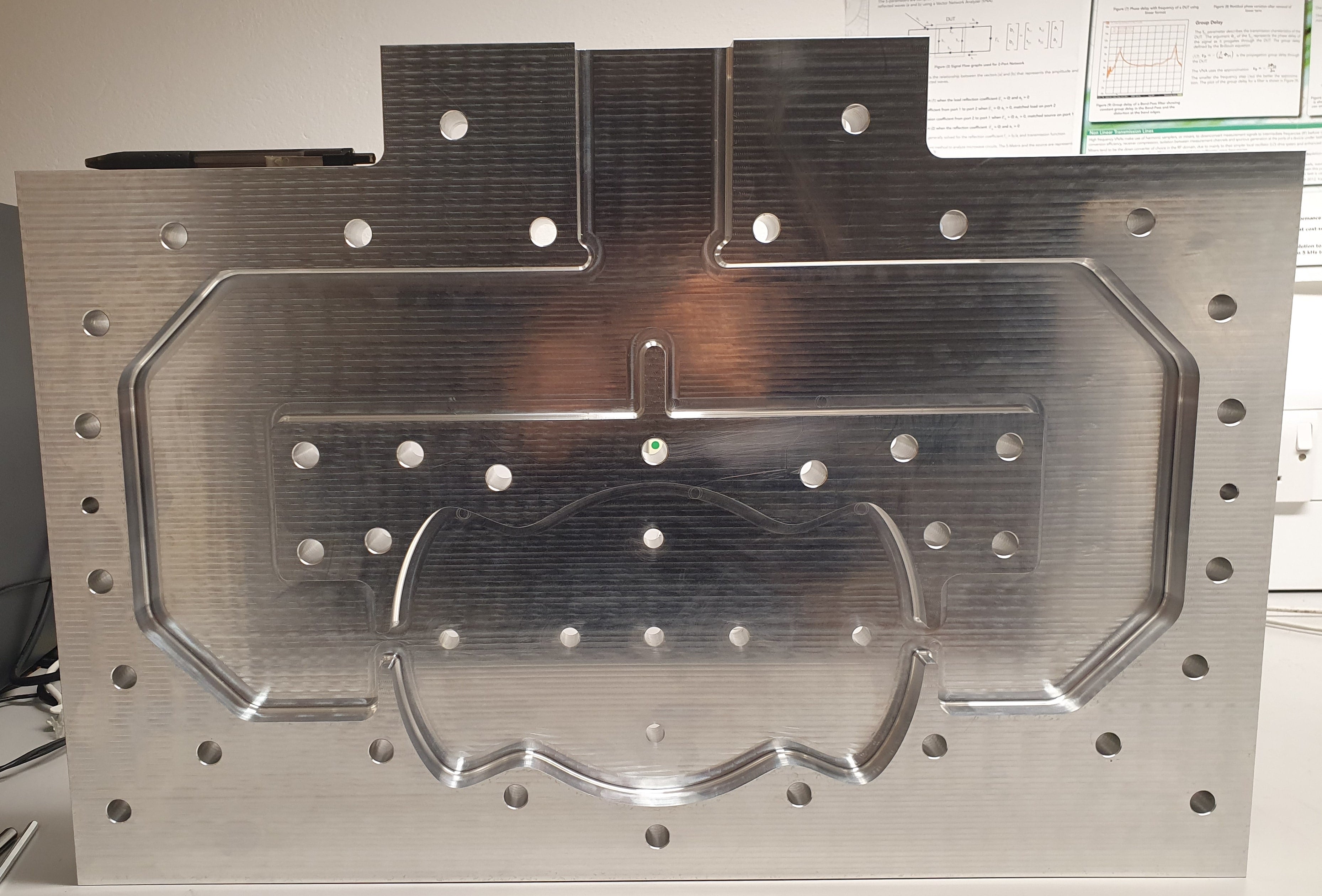}
        \caption{Half  the RF cavity system designed to support the TM$_{\{0,2\}20}$ mode with quadrupolar component 5.041 times greater than the monopolar component.}  \label{fig:QuadBuild}
\end{figure}

The RF cavity was designed as two identical halves to be sealed together and was milled out of aluminium. Seven 10 mm diameter holes were drilled into the RF cavity to create holes with centres at the locations of the seven poles in the TM$_{\{0,2\}20}$ mode. One half the RF cavity system is photographed in Fig. \ref{fig:QuadBuild}. 

\subsection{Characterisation of the RF Cavity System}

\begin{figure}[h]
    \centering
       \includegraphics[width=0.48\textwidth]{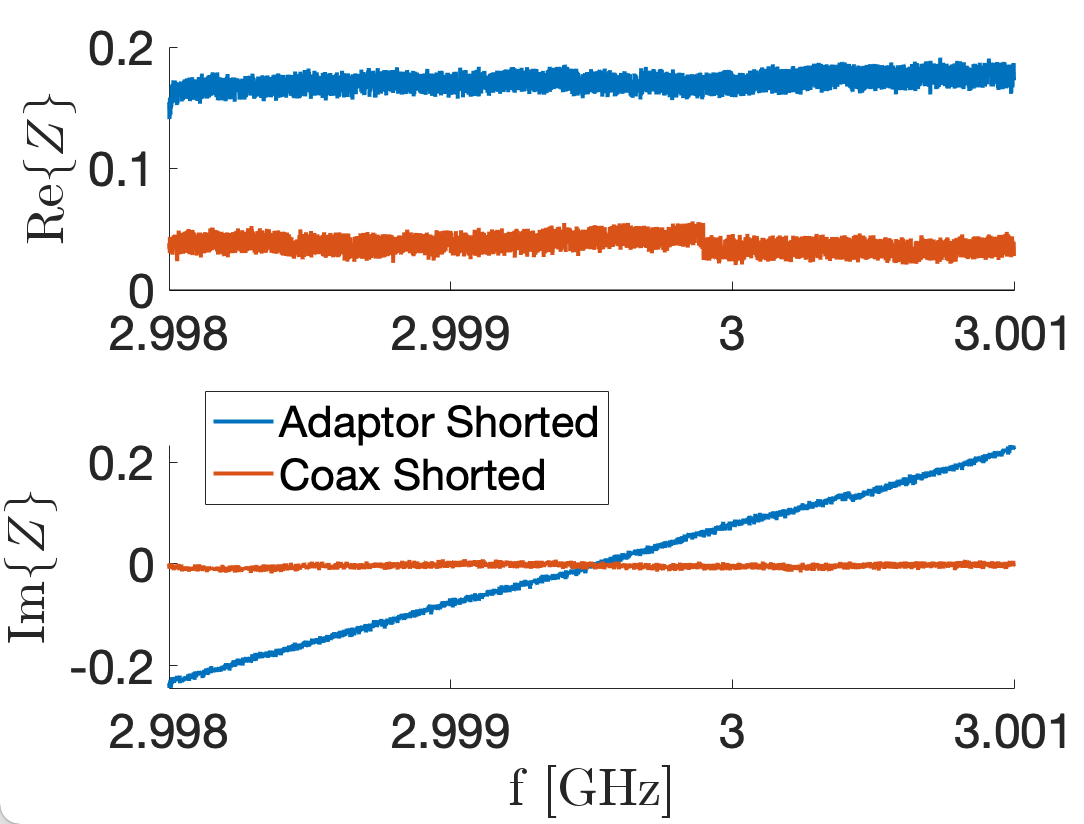}
        \caption{Impedance measurements of the calibrated coaxial cable shorted, and with the waveguide adaptor attached to the calibrated coax and shorted by clamping an aluminium plate to it.} 
        \label{fig:WaveguideMeas}
\end{figure}

Measurements of the prototype were taken using a Vector Network Analyser (VNA) connected to a matched waveguide adaptor secured to the top of the cavity waveguide port via coaxial cable. The apparatus was calibrated at the end of the coax, which excluded the matched waveguide adaptor from calibration. To ensure this does not impact results, two cases are compared in Fig. \ref{fig:WaveguideMeas}: the frequency response of shorting the calibrated coax, and of attaching the waveguide adaptor to the calibrated coax and clamping an aluminium plate to it. The maximum magnitude in the difference of the impedance measurements between the two cases is 0.2 $\Omega$. This is of the order of 100 times smaller than the changes in impedance measured in the bead pull testing and that used for calculating Q-factors. In addition to this, measurements do not give any evidence for resonances in the waveguide adaptor and the external Q-factor does not change significantly between simulations without the waveguide adaptor and measurements with the waveguide adaptor. We thus conclude that the waveguide adaptor's effect on measurements can be neglected.

The resonant frequency of the RF cavity system, $f_0$, and its intrinsic, $Q_0$, external, $Q_e$ and loaded, $Q_L$, Q-factors can be calculated from measurement of the variation in the S-parameter $S_{11}$ with driving frequency and using:
\begin{equation}
    \beta = \frac{Q_0}{Q_e} = \frac{Q_0-Q_L}{Q_L} = \frac{1-S_{11}(f=f_0)}{1+S_{11}(f=f_0)} ,
\end{equation}
where $f$ is the driving frequency and $\beta$ is the coupling coefficient. It should be noted that at resonance $S_{11}$ is purely real.

\begin{figure}[h]
    \centering
       \includegraphics[width=0.48\textwidth]{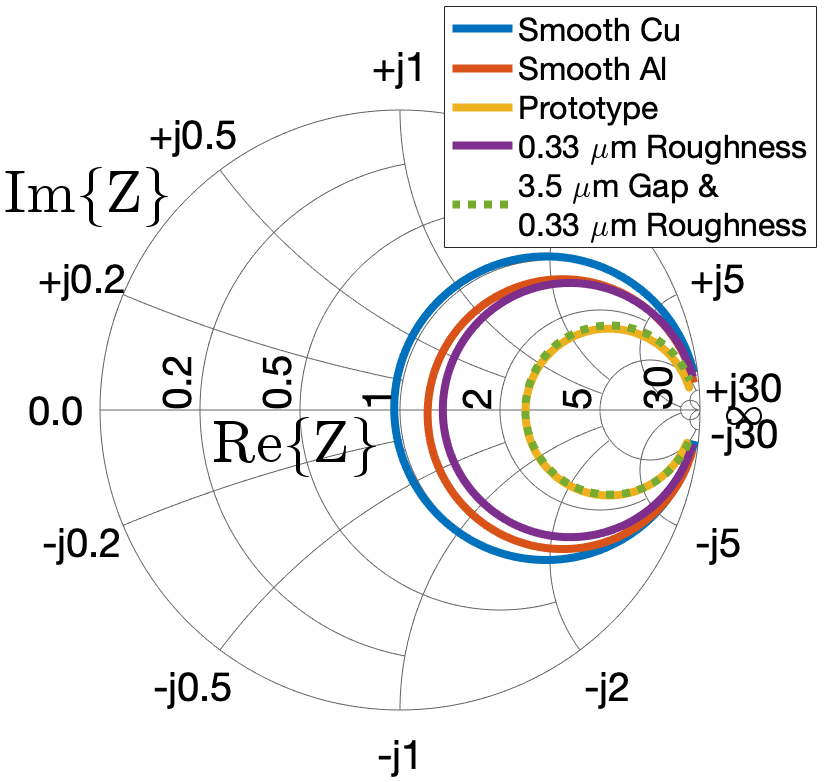}
        \caption{Smith chart showing the variation in simulation output and experimental  measurements of $S_{11}$ in the range $f_0\pm 1$ MHz.} 
        \label{fig:CavityMeasurements}
\end{figure}

Figure \ref{fig:CavityMeasurements} is a Smith chart that shows the measured variation in $S_{11}$ for driving frequencies in the range $f_0\pm 1$ MHz for five different simulation and experimental measurements. The Smith chart maps the impedance, $Z$, onto the $S_{11}$ plane and provides a method for visualising $\beta$ \cite{SmithRef}. 

The blue line of Fig. \ref{fig:CavityMeasurements} represents the simulation output of $S_{11}$ of the RF cavity system modelled as perfectly smooth copper; the electric field from this simulation was shown in Fig. \ref{fig:QuadBuildCST}. From this data, we calculate $f_0$ = 2.99914 GHz, $Q_0$ = 20830, $Q_e$ = 20050 and so the system is critically coupled as $\beta \simeq 1$. The RF cavity coupling ports were optimised for copper, while the prototype was milled out of aluminium to reduce cost. The red line in Fig. \ref{fig:CavityMeasurements} shows the simulation output for perfectly smooth aluminium and the impact of the choice of material on $\beta$: we calculate $f_0$ = 2.99917 GHz, $Q_0$ = 16880 and $\beta = 0.831$. The yellow line  represents the measurement of the prototype, and we calculate that $f_0 = 2.99914$ GHz, $Q_0 = 9120$ and $\beta = 0.410$. The prototype is thus significantly undercoupled compared to the perfectly smooth simulation. 

This discrepancy in $\beta$ is likely due to two effects. Firstly, the surface roughness due to machining of the aluminium which leads to increased resistive losses  \cite{SurfaceRoughnessRef}. The RMS surface roughness was measured with a white light interferometer to be  0.33 $\mu$m. The purple line shows the simulation output when this level of roughness is included - $f_0$ = 2.99955 GHz, $Q_0$ = 15150 and $\beta = 0.750$. The second effect is RF leakage due to an imperfect seal between the two halves of the prototype. This can be modelled by uniformly separating the two cavity halves: the  green dashed line shows that a 0.33 $\mu$m surface roughness and 3.5 $\mu$m uniform air gap between the two halves replicates the coupling coefficient of experimental measurements - $f_0$ = 2.99955 GHz, $Q_0$ = 8490 and $\beta = 0.410$.

In reality the separation between the two halves of the cavity will not be uniform, however, this would be difficult to model. Simulations show the prototype is particularly sensitive to gaps near the coupling between the waveguide and the cavity and a separation on the order of microns is feasible given the distribution of the bolts used for tightening and the presence of deeper scratches on the aluminium surfaces that are bolted together.

\subsection{Experimental Method}
Despite the RF cavity system being significantly undercoupled, the electric field of the RF cavity can be determined using the bead pull technique  \cite{HallThesis,BeadPull1,BeadPull2}. This method uses the result from cavity perturbation theory \cite{Slaters} that when a dielectric bead, small enough such that the EM fields are approximately constant over it, is introduced into an RF cavity at a position $(r,\theta,z)$: 
\begin{equation} \label{eq:BeadShift}
    \Delta f \propto |E_0(r,\theta,z)|^2,
\end{equation}
where $\Delta f$ is the change in resonant frequency caused by introducing the perturbing object and $E_0(r,\theta,z)$ is the magnitude of the electric field at that position. In our experiments, we used a spherical, plastic bead with a 2.2 mm diameter as the perturbing object.

\begin{figure}[h]
    \centering
       \includegraphics[width=0.48\textwidth]{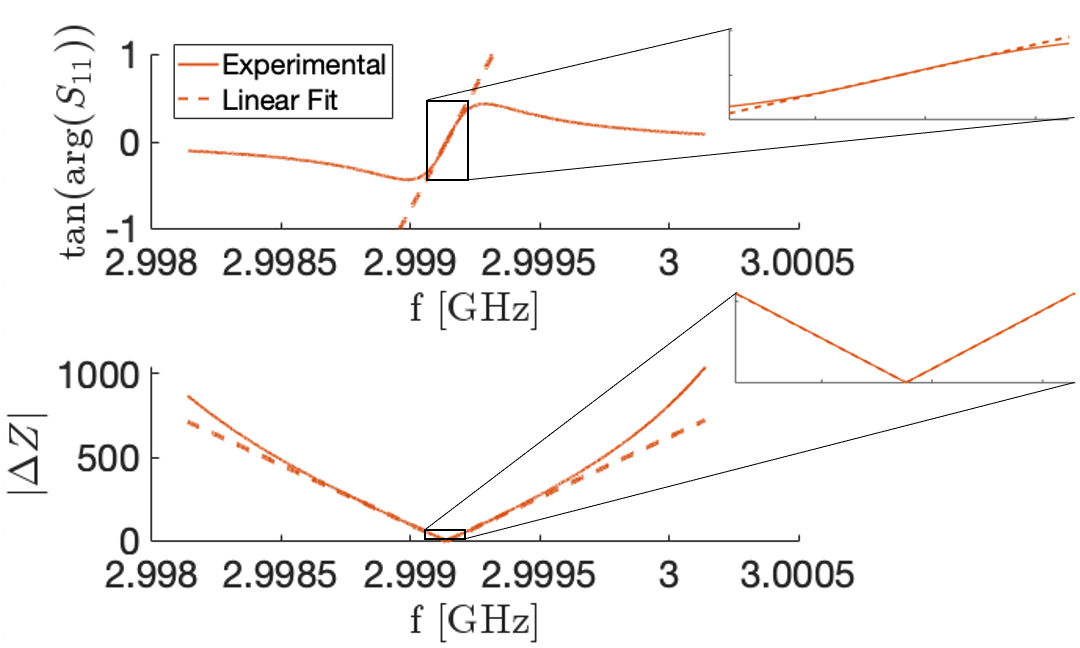}
        \caption{Measured variation in $\tan{(\arg{(S_{11})})}$ (top) and $|\Delta Z|$ (bottom) with driving frequency. The insets zoom in on the region contained inside the black rectangle.} 
        \label{fig:FreqSweepLin}
\end{figure}

The VNA cannot be used to measure directly the change in resonant frequency of the RF cavity caused by the perturbing object at a high enough sample rate. An indirect measurement is therefore required. Two possible parameters to use for this indirect measurement are the change in phase of $S_{11}$  or $|\Delta Z|$, where $|\Delta Z| = |Z - Z_\textrm{res}^\textrm{unpert}|$ and $Z^\textrm{unpert}_\textrm{res}$ is the impedance  of the unperturbed cavity at the resonant frequency. Figure \ref{fig:FreqSweepLin} shows experimental measurements of the variation in $\tan{(\arg{(S_{11})})}$ (top) and $|\Delta Z|$ (bottom) of the unperturbed cavity with $f$. The dashed lines are linear fits between the respective parameter and $f$, fitted to and extrapolated from small $\Delta f$.  The 2.2 mm diameter perturbing bead caused a maximum change in resonant frequency of $|\Delta f|$ = 0.079 MHz and the inset of Fig. \ref{fig:FreqSweepLin} zooms in on this region. It shows $|\Delta Z|$ varies linearly with $f$ to within 0.3 \%, compared to $\tan{(\arg{(S_{11})})}$ which diverges by up to 20 \%. 

We can therefore determine the electric field at a position $(r,\theta,z)$ in the RF cavity by measuring, at a fixed driving frequency, the change in impedance  caused by the bead at that position and using the relationship:
\begin{equation} \label{eq:EBead}
    |E_0(r,\theta,z)| \propto \sqrt{|\Delta Z|}.
\end{equation}

To carry out a bead pull measurement, we threaded kevlar wire through one of the holes before centering it using a digital motor system. The  2.2 mm diameter, dielectric bead knotted onto the wire was then pulled through the cavity at a speed of 3 mm/s using a longitudinal motor. The VNA was set up to continuously measure the impedance of the RF cavity at the driving frequency of 2.99914 GHz and sample rate of over 50 points per second. The change in impedance of the RF cavity was measured for the bead being pulled through all 7 holes in both longitudinal directions.

\subsection{Experimental Results}

The data was exported and analysed in MATLAB \cite{MATLAB}. Data collected when the bead was outside the beam pipes, that is $|z|>38.1$ mm where $z$ is the longitudinal position relative to the centre of the RF cavity, was removed from the data set. A line of best fit between the first and last 50 data points was then used to remove the small linear drift in impedance over time we observed.

\begin{figure}[h]
    \centering
       \includegraphics[width=0.48\textwidth]{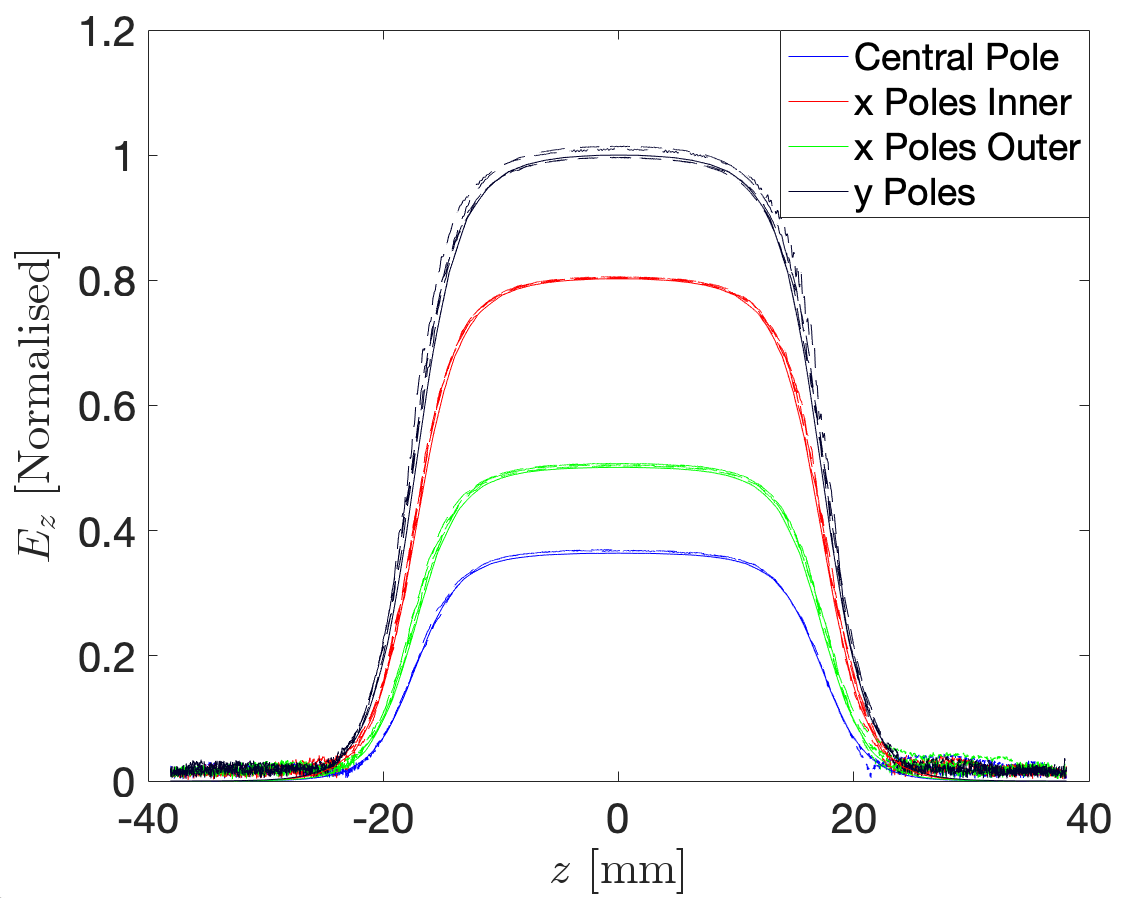}
        \caption{Comparison of bead pull measurements of the electric field compared to simulation.}  \label{fig:QuadBeadResults}
\end{figure}

Figure \ref{fig:QuadBeadResults} compares the longitudinal electric field measurement along each of the seven poles from a CST simulation of the cavity system to the experimental bead pull measurements, where we assume $|E_0|=|E_z|$ for the  measurements. Simulation data is plotted as full lines and experimental data as dashed. The different colours correspond to measurements along the different poles: blue denotes the single central pole, red the two inner horizontal ($x$) poles, green the two outer horizontal ($x$) poles, and black the two vertical ($y$) poles. To normalise the simulation data, all data points were divided by the maximum electric field in the $y$ pole. To normalise the experimental data, all data points were divided by the ratio of the mean value of the median experimental electric field measurements between $|z|<1$ mm along each pole to the mean value of the maximum simulated electric field measurements along each pole. The similarity between experimental and simulation results indicate that the cavity is symmetric in the horizontal, vertical, and longitudinal directions, as expected.

\begin{table}[b]
\caption{\label{tab:BeadPullMeas}
Comparison between simulation and experimental measurements of $E_z$ through the different beam pipe locations.}
\begin{ruledtabular}
\begin{tabular}{ccc}
Bead Pull Along & Simulation & Experimental \\
\hline
y Poles  & 1.000 & 1.004$\pm 0.008$ \\ 
x Poles Inner  & 0.802 & 0.805$\pm 0.001$ \\
x Poles Outer  & 0.501 & 0.506$\pm 0.002$ \\
Central Pole  & 0.364 & 0.368$\pm 0.001$  \\
\end{tabular}
\end{ruledtabular}
\end{table}

Table \ref{tab:BeadPullMeas} presents a quantitative comparison between the experimental and simulation results with the mean of the electric field measurement between $|z| < 1$ mm shown alongside the standard deviation of the measurements in that range. $E_z$ is expected to be flat in this range with a standard deviation below three decimal places and the results show the experimental data is consistent with simulation. It should be noted that errors arising from inaccuracies aligning the wire to the centre of the holes or from divergences from the assumed linearity in Eq. \ref{eq:EBead} have not been included in this analysis.

The data collected thus support that the electric field of the RF cavity is as designed, within error. Therefore we have used successfully the presented theory, in conjunction with 3D simulations to determine the quadrupolar term introduced by the dual-port coupler, to design and measure the field of an azimuthally modulated RF cavity that supports a TM$_{\{0,2\}20}$ mode with a quadrupolar to monopolar ratio of 5.041.

\section{Conclusion} \label{sec:Conclusion}

This paper has derived the basis of the longitudinal electric fields in azimuthally modulated RF cavities and presented a systematic method for determining the cavity shape that will support TM$_{\{M\}\eta0}$ modes. Examples of the RF cavity shapes that support  TM$_{\{0,m_1\}\eta0}$ or TM$_{\{m_1,m_2\}\eta0}$ modes have been explicitly examined with rules derived for RF cavity shapes being conditional, forbidden or spiral depending upon the values of $m_1$ and $\eta$, and $m_1$, $m_2$ and $\eta$. The theoretical work has been supported by analysis of 3D simulations and experimental test of an RF cavity that supports a TM$_{\{0,2\}20}$ mode.

The results of this paper provide a basis for the development of new designs of azimuthally modulated RF cavities that contain any magnitude and order of multipole components. These designs may have useful application in particle accelerators, particularly as they do not have degenerate same-order modes and as they can introduce or cancel wanted or unwanted multipolar components to a fine precision.

Future research can use the formalism and results presented in this paper to guide the development of approximate analytic models of cavity designs and understand how the variation of certain design parameters affects the cavity performance. This could reduce the need to run computationally expensive simulations. Additionally, the paper provides a basis for determining the exact effect of the Bessel dependence of the transverse deflecting fields on beam dynamics. Future studies could also investigate the introduction of longitudinal asymmetries alongside the azimuthal asymmetries in the RF cavity with the aim of enhancing the multipolar content of a field and creating additional parameters with which to optimise  performance in aspects such as shunt impedance. 

\section{Acknowledgements}

We would like to thank Graeme Burt for many useful discussions and particularly for his assistance in designing the coupler for the RF cavity system. We would also like to thank Shinji Machida, Jean-Baptiste Lagrange, David Kelliher, Chris Rogers and Chris Prior of the ISIS Intense Beams Group for their insightful discussions and feedback, as well as  Manjit Dosanjh who has contributed greatly to the direction of the project and regarding the application of discussed work. This study and the experimental work has been funded by the Royal Society under RS$\backslash$PhD$\backslash$181200 and RGF$\backslash$EA$\backslash$18023, and the CI Core Grant ST/P002056/1 has funded travel.

\pagebreak

\appendix

\section{\label{sec:AppForbiddenSpiral}Forbidden and Spiral Modes}

Here we discuss why certain TM$_{\{m_1,m_2\}\eta0}$ modes are forbidden, that is $r^{(\eta)}(\theta)$ = 0 for some $\theta$ value, or spiral, that is $r^{(\eta)}(0) \neq r^{(\eta)}(2\pi)$. 

As our first step, we prove that these $r_0^{\eta}(\theta)$ solutions are not necessarily bounded within an interval $r = [j_{m_2(x-1)}, j_{m_2x}]/k$ ($x \in \mathbb{Z}$). These solutions must solve the boundary condition in Eq. \ref{eq:TwoMultipole} that we replicate here:
\begin{multline}\label{eq:AppTwo}
    J_{m_1}\left(k r_0^{(\eta)}(\theta)\right)\tilde{g}_{m_1}\cos{m_1\theta} +\\
    J_{m_2}\left(k r_0^{(\eta)}(\theta)\right)\tilde{g}_{m_2}\cos{(m_2\theta-\phi_{m_2})}= 0.
\end{multline}
Firstly, consider Eq. \ref{eq:AppTwo} for $\theta=0$: it follows that $r_0^{(\eta)}(0) \neq j_{m_2y}/k$ ($y \in \mathbb{Z}$). This means that all solutions $r_0^{(\eta)}(0)$ lie within an interval $r = [j_{m_2(x-1)}, j_{m_2x}]/k$ and not at the limits, or edges, of the interval. Next we consider the unique angles in the range $[0, 2\pi]$, $\theta_q = (2q+1)\pi/2m_1$ ($q = 0, 1,\cdots, \lfloor2m_1-1/2\rfloor$), that eliminate the cosine term of order $m_1$ such that Eq. \ref{eq:AppTwo} becomes:
\begin{equation} \label{eq:AppPossibility}
    J_{m_2}\left(k r_0^{(\eta)}(\theta_q)\right)\tilde{g}_{m_2}\cos{\left(\frac{m_2}{m_1}(2q+1)\frac{\pi}{2}-\phi_{m_2}\right)} = 0.
\end{equation}

This has two possible solutions whereby either the Bessel term of order $m_2$ is zero or the cosine term is zero. We will firstly consider the cosine term to be non-zero, which forces:
\begin{equation} \label{eq:IntCond}
    k r_0^{(\eta)}(\theta_q) = j_{m_2y},
\end{equation}
to satisfy Eq. \ref{eq:AppPossibility}. Equation \ref{eq:IntCond} thus states that if $\cos{(m_2\theta_q-\phi_{m_2})}\neq0$, $r_0^{(\eta)}(\theta_q)$ must lie at the limit of an interval $r = [j_{m_2(x-1)}, j_{m_2x}]/k$.  Therefore the solution will cross the upper limit of the interval if the gradient is positive at $\theta_q$, whereas it will cross the lower limit if it is negative. Differentiating Eq. \ref{eq:AppTwo} we find:
\begin{widetext}
\begin{equation} \label{eq:GenDiffCond}
    \frac{dr_0^{(\eta)}(\theta)}{d\theta} = \frac{1}{k}\frac{J_{m_1}(kr_0^{(\eta)}(\theta))\tilde{g}_{m_1}m_1\sin{m_1\theta}+J_{m_2}(kr_0^{(\eta)}(\theta))\tilde{g}_{m_2}m_2\sin{(m_2\theta-\phi_{m_2})}}
    {J_{m_1}'(kr_0^{(\eta)}(\theta))\tilde{g}_{m_1}\cos{m_1\theta}+J_{m_2}'(kr_0^{(\eta)}(\theta))\tilde{g}_{m_2}\cos{(m_2\theta-\phi_{m_2})}},
\end{equation}
\end{widetext}
which if $\cos{(m_2\theta_q-\phi_{m_2})}\neq0$ becomes:
\begin{equation} \label{eq:AppGradient}
    \frac{dr_0^{(\eta)}(\theta)}{d\theta}\bigg|_{\theta_q} = \frac{m_1\tilde{g}_{m_1}}{k\tilde{g}_{m_2}}\frac{J_{m_1}(j_{m_2y})(-1)^{q}}
    {J_{m_2}'(j_{m_2y})\cos{m_2\theta_q}} \neq 0.
\end{equation}
Equation \ref{eq:AppGradient} can be numerically evaluated to determine the sign of the gradient at all angles $\theta_q$ at the limits of each interval $j_{m_2y}$. We conclude that if the number of times the gradient is positive does not equal the number of times the gradient is negative  (that is the number of times  the upper limit of an interval is crossed does not equal the total number of times the lower limit of an interval is crossed) then $r^{(\eta)}(0)$ will lie in a different interval to $r^{(\eta)}(2\pi)$. This means $r^{(\eta)}(0)\neq r^{(\eta)}(2\pi)$ and the solution and mode are spiral. If, however, the number of times the upper limit is crossed equals the number of times the lower limit is crossed, then the solution will lie in the interval $r = [j_{m_2(x-\zeta_1)}, j_{m_2(x+\zeta_2)}]/k$, where the $\zeta_i$ are numerically calculated integers. If $\zeta_1\geq x$ then the corresponding TM$_{\{m_1,m_2\}\eta0}$ mode will be forbidden as it will have $r^{(\eta)}(\theta) = 0$ for some $\theta$.

\begin{table}[h!]
\caption{\label{tab:ForbiddenModes}
The $\eta$ values of the first non-forbidden TM$_{\{m_1,m_2\}\eta 0}$ mode shown for the case $\phi_{m_2}\neq\phi_{m_2}^{(q)}$ (left of comma) and $\phi_{m_2}=\phi_{m_2}^{(q)}$ (right of comma). $\nexists$ denotes a spiral mode.}
\begin{ruledtabular}
\begin{tabular}{cccccccccc}
\diagbox{$m_2$}{$m_1$} & 1 & 2 & 3 & 4 & 5 & 6 & 7 & 8 & 9 \\
\hline
2 & 2, 1 & - & - & - & - & - & - & - & - \\
 3 & $\nexists$, 1 & 3, 2 & - & - & - & - & - & - & - \\
 4 & 2, 1 & 2, 1 & 4, 3 & - & - & - & - & - & - \\
 5 & $\nexists$, 1 & 3, 2 & $\nexists$, 2 & 5, 4 & - & - & - & - & - \\
 6 & 2, 1 & $\nexists$, 1 & 2, 1 & 3, 2 & 6, 5 & - & - & - & - \\
 7 & $\nexists$, 1 & 3, 2 & $\nexists$, 2 & 3, 2 & $\nexists$, 3 & 7, 6 & - & - & - \\
 8 & 2, 1 & 2, 1 & 4, 3 & 2, 2 & 4, 3 & 4, 3 & 8, 7 & - & - \\
 9 & $\nexists$, 1 & 3, 2 & $\nexists$, 1 & 3, 2 & $\nexists$, 2 & 3, 2 & $\nexists$, 4 & 9, 8 & - \\
 10 & 2, 1 & $\nexists$, 1 & 4, 3 & 3, 2 & 2, 1 & $\nexists$, 2 & 4, 3 & 5, 4 & 10, 9 \\
\end{tabular}
\end{ruledtabular}
\end{table}

We now come back to Eq. \ref{eq:AppPossibility} and allow the cosine term to be zero. This requires the phase to take one of the following $q$-dependent values:
\begin{equation} \label{eq:AppPhiVal}
    \phi_{m_2} = \phi_{m_2}^{(q)} = \frac{\pi}{2}\left(\frac{m_2}{m_1}(2q+1)-(2x+1)\right).
\end{equation}
The effect of setting the phase term equal to $\phi_{m_2}^{(q)}$ is that at the corresponding $\theta_q$ angle, $r_0^{(\eta)}(\theta_q)$ does not lie on the limit of an interval and thus the interval is not crossed.

\begin{table}[h!]
\caption{\label{tab:PhiVals}
The value of $\phi^{(q)}_{m_2}$ for TM$_{\{m_1,m_2\}\eta 0}$ modes. O denotes odd integer and N any integer.} 
\begin{ruledtabular}
\begin{tabular}{cccccccccc}
\diagbox{$m_2$}{$m_1$} & 1 & 2 & 3 & 4 & 5 & 6 & 7 & 8 & 9 \\
\hline
2 & O$\frac{\pi}{2}$ & - & - & - & - & - & - & - & -  \\
 3 & N$\pi$ & O$\frac{\pi}{4}$ & - & - & - & - & - & - & - \\
 4 & O$\frac{\pi}{2}$ & O$\frac{\pi}{2}$ & O$\frac{\pi}{6}$ & - & - & - & - & - & - \\
 5 & N$\pi$ & O$\frac{\pi}{4}$ & N$\frac{\pi}{3}$ & O$\frac{\pi}{8}$ & -  & - & - & - & - \\
 6 & O$\frac{\pi}{2}$ & N$\pi$ & O$\frac{\pi}{2}$ & O$\frac{\pi}{4}$ & O$\frac{\pi}{10}$  & - & - & - & -   \\
 7 & N$\pi$ & O$\frac{\pi}{4}$ & N$\frac{\pi}{3}$ & O$\frac{\pi}{8}$ & N$\frac{\pi}{5}$ & O$\frac{\pi}{12}$ & - & - & - \\
 8 & O$\frac{\pi}{2}$ & O$\frac{\pi}{2}$ & O$\frac{\pi}{6}$ & O$\frac{\pi}{2}$ & O$\frac{\pi}{10}$ & O$\frac{\pi}{6}$ & O$\frac{\pi}{14}$ & - & -\\
 9 & N$\pi$ & O$\frac{\pi}{4}$ & N$\pi$ & O$\frac{\pi}{8}$ & N$\frac{\pi}{5}$  & O$\frac{\pi}{4}$ & N$\frac{\pi}{7}$ & O$\frac{\pi}{16}$ & - \\
 10 & O$\frac{\pi}{2}$ & N$\pi$ & O$\frac{\pi}{6}$ & O$\frac{\pi}{4}$ &  O$\frac{\pi}{2}$ & N$\frac{\pi}{3}$ & O$\frac{\pi}{14}$ & O$\frac{\pi}{8}$ & O$\frac{\pi}{18}$ \\
\end{tabular}
\end{ruledtabular}
\end{table}

The work discussed here is brought together in Table \ref{tab:ForbiddenModes} which, for given $m_1$ and $m_2$ values between 1 and 10, displays the first non-forbidden and non-spiral mode for $\phi_{m_2}\neq\phi^{(q)}_{m_2}$ (left of comma) and $\phi_{m_2}=\phi^{(q)}_{m_2}$ (right of comma). These were numerically calculated using Eq. \ref{eq:AppGradient}. The values of $\phi^{(q)}$ for given $m_1$ and $m_2$, as calculated by Eq. \ref{eq:AppPhiVal}, are displayed in Table \ref{tab:PhiVals}.

\end{document}